\documentclass[oldversion]{aa}
\usepackage{graphicx}
\usepackage{txfonts}
\def\simless{\mathbin{\lower 3pt\hbox
{$\rlap{\raise 5pt\hbox{$\char'074$}}\mathchar"7218$}}}   %< or of order
\def\simmore{\mathbin{\lower 3pt\hbox
{$\rlap{\raise 5pt\hbox{$\char'076$}}\mathchar"7218$}}}   %> or of order
\newcommand{\be}{\begin{equation}}
\newcommand{\ee}{\end{equation}}
\newcommand{\lb}[0] { \left( }
\newcommand{\rb}[0] { \right) }

\newcommand{\beqs} { \begin{eqnarray} }
\newcommand{\eeqs} { \end{eqnarray} }

\newcommand{\ep}[0] { \epsilon }

\newcommand{\eqref}[1]{(\ref{#1})}

\newcommand{\tmax}{ {\mathrm{max}} }

\newcommand{\ttot}{ {\mathrm{tot}} }

\newcommand{\EE}[2]{#1 \times 10^{#2}}

\newcommand{\trel}{ {\mathrm{rel}} }

\newcommand{\tinel}{ {\mathrm{inel}} } 
\newcommand{\tel}{ {\mathrm{el}} } 

\newcommand{\np}{$np$ }

\newcommand{\tFB}{ {\mathrm{FB}} } 
\newcommand{\tAC}{ {\mathrm{AC}} }

\begin{document}
\title{Neutron-rich gamma-ray burst flows: dynamics and particle creation in neutron -- proton collisions}
\titlerunning{Neutron-rich GRB flows}
\author{Hylke B. J. Koers\inst{1,2} \and Dimitrios Giannios\inst{3}}
\authorrunning{Hylke Koers \& Dimitrios Giannios}
\institute{Nikhef, PO Box 41882, 1009 DB Amsterdam, The Netherlands \and  University of Amsterdam, Amsterdam, 
The Netherlands \and Max Planck Institute for Astrophysics, PO Box 1317, D-85741 Garching, Germany}

\offprints{hkoers@nikhef.nl}
\date{Received / Accepted}

\abstract
{We consider gamma-ray burst outflows with a substantial neutron component that are 
either dominated by thermal energy (fireballs) or by magnetic energy.  In the latter case,
we focus on the recently introduced `AC' model which relies on magnetic reconnection to accelerate
the flow and power the prompt emission. For both the fireball and the AC model, we investigate
the dynamical importance of neutrons on the
outflow. We study particle creation in inelastic neutron -- proton
collisions and find that in both models the resulting neutrino emission
is too weak to be detectable. The inelastic collisions also 
produce $\gamma$-rays, which create pairs in interactions with  soft photons
carried with  the flow.
In magnetically driven outflows, the energy of these pairs is radiated away as synchrotron emission. The
 bulk of the emission takes place at a few hundred keV, which makes it difficult to disentangle this signal
from the prompt emission.
In fireballs, however, pair cascading leads to the emission of $\gamma$-rays with observer energy in the
range of 2 - 20 GeV and a fluence well above
the GLAST threshold.  
Therefore this emission can be a useful diagnostic of the nature of the outflow.
}

\maketitle

\section{Introduction} 
\label{intro}

In recent years there has been significant progress in our understanding of $\gamma$-ray bursts
(GRBs). The observational connection between supernovae and GRBs and studies of GRB host
galaxies provide compelling evidence for a connection between long GRBs and the death of massive stars
(Van Paradijs et al. 2000; Woosley \& Bloom 2006). 
The general scenario for long GRBs (for recent reviews, see Piran (2004);
M\'esz\'aros (2006)) starts with core collapse of the massive star leading to the
formation of a black hole
surrounded by an accretion disk. The black hole -- accretion disk system powers a developing
outflow along the rotational axis, which accelerates to a bulk Lorentz factor of a few hundred, 
transferring its energy to the baryons contained in the flow. Dissipation of 
energy in the outflow leads to the
prompt $\gamma$-ray emission while  the interaction of the outflow with the
external medium results in the afterglow. 

The nature of the relativistic outflow is currently one of the most important
open questions regarding GRBs. The high Lorentz factor, required to match the inferred energy density of the source
and the observed non-thermal character of the emission  (the compactness
problem; see e.g. Piran (2004)), implies that 
the ratio of energy to rest mass of the flow must be very high. 
In the widely used fireball model (Cavallo \& Rees 1978; Goodman 1986;
Paczy\'nski 1986) the outflow is a 
photon-electron-positron plasma that is
dominated by thermal energy and has a small baryonic load.
Alternatively, the energy of the outflow may initially be
dominated by Poynting flux (Usov 1992). Such  outflows occur naturally when a magnetized
accretion disk surrounds a black hole (Thompson 1994; M\'esz\'aros \&
Rees 1997; Spruit et al. 2001;  Van Putten \& Ostriker 2001; Vlahakis \& K\"onigl
2001; Drenkhahn \& Spruit 2002; Lyutikov \& Blandford 2003; Lyutikov 2006;
Uzdensky \& MacFadyen 2006).

Neutrinos and $\gamma$-rays may be useful probes to differentiate between
fireballs and Poynting-flux dominated (PFD) outflows.
The internal shocks that are believed to accelerate electrons in the
fireball model will also accelerate protons to very high energies, giving
rise to neutrinos with energy $\gtrsim$$100$ TeV through photopion production
(Waxman \& Bahcall 1997). In the absence of a mechanism to accelerate protons
to very high energies these neutrinos are not expected in PFD outflows.
In this paper we  consider \mbox{neutron -- proton ($np$)} collisions in neutron-rich flows
and address the question whether neutrinos and $\gamma$-rays created in
these hadronic interactions can also be used to probe the nature of GRB outflows.

GRB outflows are expected to be neutron-rich. In GRB central engines,
the competition of positron capture on neutrons and electron capture on protons
favours a neutron-rich environment (Beloborodov 2003b;  Pruet et al. 2003;
Chen \& Beloborodov 2007).
Nucleosynthesizing interactions reduce the number of free neutrons in the
outflow, but a significant amount of neutrons remains in the flow until neutron decay
becomes important (Beloborodov 2003b; Inoue et al. 2003). 
Deep in the  outflow protons and neutrons are strongly coupled  through
nuclear scattering and behave as a single fluid that accelerates to high Lorentz factors.
With increasing distance from the central engine
the densities decrease until neutrons decouple and enter the coasting phase.
Protons, being electromagnetically coupled to the flow, may be accelerated further.
When the relative velocity between neutrons and protons is sufficiently high, 
 inelastic \np collisions are possible and lead to pion creation. The
pions decay into $\gamma$-rays and neutrinos with observer energies in the 
$\sim$$10-100$ GeV
range. This mechanism has
been investigated for fireballs (Derishev et al. 1999a;
Bahcall \& M\'esz\'aros 2000; 
M\'esz\'aros \& Rees 2000; Belyanin et al. 2003; Razzaque \& M\'esz\'aros
2006) but, to the best of our knowledge, not for PFD flows.

The creation of secondary particles in inelastic \np collisions can potentially be used
to identify a substantial neutron component in GRB flows. Other
ways to identify such a component that have been suggested in the
literature are through signatures 
in the early afterglow of GRBs 
(Derishev et al. 1999b, Beloborodov 2003a; Fan, Zhang \& Wei 2005),
ultraviolet flashes generated in internal shocks in neutron-rich flows (Fan \& Wei 2004),
and observational signatures of a two-component jet that may be associated with
neutron-rich MHD flows (Vlahakis et al. 2003; Peng et al. 2005).

In this work we consider the `AC' model as a specific model for PFD outflows.
In this model the magnetic field configuration is similar to that produced by an inclined
rotator (Coroniti 1990;  Lyubarsky \& Kirk 2001) with field lines changing polarity on a scale
$\lambda \simeq 2 \pi c / \Omega$, where $\Omega$ denotes the angular frequency of the rotator.
This model was recently discussed in connection to GRBs in a series of papers
(Spruit et al. 2001;
Drenkhahn 2002; Drenkhahn \& Spruit 2002; Giannios \& Spruit 2005; Giannios 2006),
where it was found that dissipation of the electromagnetic energy by magnetic reconnection can account 
for both the bulk acceleration of the flow and for the prompt emission.

The dynamics of fireballs and of outflows in the AC model are distinctively different.
Fireballs are driven by radiation pressure and exhibit a 
period of rapid acceleration in which the Lorentz factor $\Gamma \propto r$,
where $r$ denotes the distance from 
the central engine (Paczy\'nski 1986). The flow saturates either when there is no more energy available
to further accelerate the baryons or when radiation and matter decouple at the
Thomson photosphere.
An analysis of the dynamics of neutron-rich fireballs was recently 
presented by Rossi et al. (2006). 
The dynamics of neutrons in MHD flows was considered previously by Vlahakis et al. (2003)
in the context of a different model for the outflow (Vlahakis \& K\"onigl 2003) than the
AC model considered here.
In the AC model, the acceleration of the flow is quite gradual and can be approximated with
$\Gamma \propto r^{1/3}$ (Drenkhahn 2002). Since acceleration of the flow 
is driven by magnetic forces, the flow can saturate far beyond the photosphere.
It is expected that the difference in dynamics affects the number and the energy
of secondary particles created in \np collisions. Furthermore,
the presence of a strong magnetic field can affect the interaction of secondary
particles  with the flow.

Motivated by the fact that neutrinos and $\gamma$-rays from inelastic \np
collisions could provide an indication 
about the nature of GRB outflows,
we consider in this paper both fireballs and AC flows with a substantial
neutron component. We investigate 
the dynamics of these flows and the creation of $\gamma$-rays and neutrinos in inelastic \np collisions.
In order to give an accurate comparison between
the fireball model and the AC model, we consider both models here. Furthermore we
use accurate fitting formulae for both the total and inelastic \np cross sections,
which has an important effect on the calculated fluences of secondary particles.

This paper is organized as follows. In section~\ref{dynamics} we discuss
the dynamical behavior of fireballs and of GRB outflows described by the AC model.
In section~\ref{sec:particlecreation} we consider
particle creation in inelastic \np collisions. We discuss here the parameter space in which the
mechanism is operational and we compute the fluences and energies of secondary
neutrinos and $\gamma$-rays. Detection prospects are discussed in section~\ref{sec:detection} and
conclusions are presented in section~\ref{sec:conclusions}.

\section{Dynamics of neutron-rich GRB flows}
\label{dynamics}
Deep in the flow neutrons are strongly coupled to protons through
elastic collisions, so that the two fluids behave as a single one.
 This \np fluid  is accelerated by conversion
of thermal energy into kinetic energy in the fireball model
and of magnetic energy into kinetic energy in the reconnection model.
When the dynamical time of the flow becomes shorter than the 
\np collision time, the two fluids decouple and the neutrons
enter the coasting phase.
Provided that the flow has not already reached its terminal bulk Lorentz factor,
the protons keep accelerating above the \np  decoupling radius, 
 which results in relative motion of the two fluids.

The analysis of the effect of a neutron component on the dynamics 
is made separately for the fireball and the reconnection model
for the various stages of their evolution.
Since  the treatment of the mass flux is 
identical in both  models, it is presented first.

\subsection{Mass flux: protons and neutrons} 

For an ultrarelativistic, steady, radial flow, assumed by both models under
consideration, 
conservation of mass implies that the baryon outflow rate obeys
\be
\label{Mdot}
\dot M=\dot M_{\rm p}+\dot M_{\rm n}=4\pi r^2mc(\Gamma_{\rm p} n'_{\rm p}
+\Gamma_{\rm n}n'_{\rm n})=4\pi r^2mc\Gamma_{\rm p} n'_{\rm p}(1+\xi) \, ,
\ee
where $\Gamma_{\rm p}$ and $\Gamma_{\rm n}$ stand for the bulk Lorentz factor
of the protons and the neutrons, respectively, and $n'_{\rm p}$ and $n'_{\rm n}$
for their proper number 
densities. The masses of protons and neutrons are assumed equal $m_{\rm p}\simeq
m_{\rm n}=m$ and $\xi$ stands for the neutron-to-proton mass flux ratio:
\be
\xi\equiv \frac{\dot M_{\rm n}}{\dot M_{\rm p}}=\frac{\Gamma_{\rm n} n'_{\rm n}}{\Gamma_{\rm p} n'_{\rm p}} \, .
\label{ksi}
\ee
The ratio $\xi$ depends on the radius $r$ since free neutrons decay into
protons on a comoving timescale $\tau_{\beta}\sim900$ sec resulting in
\be
\frac{{\rm d}\dot M_{\rm n}}{{\rm d}r}=\frac{{\rm d}\dot M_{\rm
 n}}{\Gamma_{\rm n}c {\rm d} t'}=-\frac{\dot M_{\rm n}}{\Gamma_{\rm n}c \tau_{\beta}} \, ,
\label{dotMn}
 \ee
where $t'$ stands for the comoving time. Taking into account that  a proton is produced 
for every neutron that decays (i.e. $\rm d\dot M_{\rm n}/{\rm
d}r=-\rm d\dot M_{\rm p}/{\rm d}r$), eqs.~(\ref{ksi}) and (\ref{dotMn}) yield
an expression for $\xi$ as a function of radius:
\be
\frac{{\rm d}\xi}{{\rm d}r}=-\frac{\xi(1+\xi)}{\Gamma_{\rm n}c \tau_{\beta}} \, .
\label{ksir}
\ee
From eq.~(\ref{Mdot}) one can solve for the number density of protons
and neutrons as a function of radius to find that
\be
n'_{\rm p}=\frac{1}{1+\xi}\frac{\dot M}{4\pi r^2 mc\Gamma_{\rm p}} \, ,
\label{np}
\ee   
and
\be
n'_{\rm n}=\frac{\xi}{1+\xi}\frac{\dot M}{4\pi r^2mc\Gamma_{\rm n}} \, .
\label{nn}
\ee
The number density of the protons and neutrons 
is determined once their bulk Lorentz factor as a function
of radius is derived. This is the topic of the next sections.

\subsection{The fireball}
\label{sect:fb}

In the fireball model most of the energy is initially stored in the 
form of thermal energy $e$,  which is dominated by the energy density of radiation.
The luminosity $L$ of the flow is the sum of kinetic and radiation flux
(e.g., Rossi et al. 2006):
\be
L=4\pi r^2c\Big[\Gamma_{\rm p}^2(4e/3+n'_{\rm p}mc^2)+\Gamma_{\rm n}^2n'_{\rm
    n}mc^2\Big] \, .
\label{LumFB}
\ee
This expression can be rewritten as
\be
L=4\pi r^2\Gamma_{\rm p}^2cn'_{\rm p}mc^2\big(1+\xi\frac{\Gamma_{\rm
  n}}{\Gamma_{\rm p}}+x\big) \, ,
\label{Lfireball}
\ee
where we have defined $x\equiv 4e/3n'_{\rm p}mc^2$.

An important quantity for the evolution of the flow is the baryon
loading parameter $\eta\equiv L/\dot M c^2\gg 1$ where $\dot M$ (defined in
eq. (\ref{Mdot})) includes both the
 contribution of the proton and the neutron fluid. Using expressions
(\ref{Mdot}) and (\ref{Lfireball}) one derives the expression
\be
(1+\xi)\eta=\Gamma_{\rm p}(1+x)+\xi\Gamma_{\rm n} \, .
\label{eta}
\ee
Assuming that the flow starts from rest
(i.e., $\Gamma_{\rm p, 0}=\Gamma_{\rm n,0}=1$) at an initial radius $r_0$ 
and initial neutron-to-proton ratio $\xi_0$, the
initial value for $x$ is $x_0= 4e_0/3n'_{\rm p,0} mc^2=
(1+\xi_0)(\eta-1)$. 

As long as the flow is Thomson thick, radiation and particles remain coupled
and the evolution of the fireball is fully determined by the adiabatic
law\footnote{This expression does not take into account
the increase of the proton density  due to neutron decay.
The use of this expression is justified because, for the parameter space relevant for GRB flows,
there is only a negligible
fraction of neutrons that decays below the photosphere of fireballs.
Hereafter, in the Thomson thick part of the flow, we set $\xi=\xi_0$.} 
\be
e=e_0\Big(\frac{n'_{\rm p}}{n'_{\rm{p,0}}}\Big)^{4/3} \, .
\label{adiabatic}
\ee
From eqs. (\ref{np}), (\ref{eta}) and (\ref{adiabatic})  one finds for the
internal energy-to-proton rest mass ratio in the flow
\be
x=x_0\Big(\frac{n'_{\rm p}}{n'_{\rm p,0}}\Big)^{1/3}=(1+\xi_0)(\eta-1)
\Big(\frac{r_0^2}{r^2\Gamma_{\rm p}}\Big)^{1/3} \, .
\label{x}
\ee  
Differentiating eq. (\ref{eta}) with respect to radius $r$ and using
eq. (\ref{x}), one has an expression relating the bulk Lorentz factor 
of the proton and the neutron fluids in the optically thick part of the flow
(see also Rossi et al. 2006)
\be
\frac{\rm d \Gamma_{\rm p}}{\rm d r}=\frac{\Gamma_{\rm
 p}}{r}\frac{2x}{2x+3}-\frac{3\xi_0}{2x+3}\frac{\rm d \Gamma_{\rm n}}{\rm d r} \, .
\label{gammap}
\ee

For the dynamics of the neutron-rich fireball to be fully determined, one
needs to look closer at the momentum exchange between the neutron and the 
proton fluids because of \np collisions. This has been studied by
Derishev et al (1999a) and Rossi et al. (2006) who showed that when
the two fluids have a relative velocity $\beta_{\rm rel}$, there is
a drag force that accelerates the neutrons
\be
\frac{\rm d \Gamma_{\rm n}}{\rm d r}=\frac{n'_{\rm p}\sigma_{\rm
      tot}}{2}\Gamma_{\rm {rel}}^2\beta_{\rm rel}^2 \, ,
\label{gamman}
\ee
where $\Gamma_{\rm rel}\simeq (\Gamma_{\rm n}/\Gamma_{\rm p}+\Gamma_{\rm
  p}/\Gamma_{\rm n})/2$ for ultrarelativistic flows and 
the total \np scattering cross section $\sigma_{\rm
  {tot}}$ is a function of $\beta_{\rm rel}$.
  This expression accounts for the  \np interaction
and does not depend on the acceleration mechanism (thermal or magnetic)
of the flow. It can, thus, be applied to both fireballs and MHD flows. 

The \np scattering cross section depends on the relative velocity
of the two fluids.
For \np  scatterings that take place with energies below the pion 
creation threshold, the scattering cross section can with good accuracy
be taken to scale as $\propto
1/(c_1 \beta_{\rm rel}+c_2 \beta_{\rm rel}^3)$, while  it remains almost
constant for higher energies. The constants $c_1$ and $c_2$ are found by fitting  to
experimental data from Yao et al. (2006; see appendix A)
\be
\sigma_{\rm tot}=\rm{max}\Big[\frac{\bar{\sigma}}{0.19\beta_{\rm rel}+5.2\beta_{\rm
    rel}^3}, \bar{\sigma}\Big] \, ,
\label{sigmatot}
\ee
 where $\bar{\sigma}\approx 4\times 10^{-26}$ cm$^2$.
Our fitting formulae for $\sigma_{\rm tot}$ are more accurate than
the expressions used by Rossi et al. (2006),  where the total \np
scattering cross section is substantially underestimated for
 $\beta_{\rm rel}\Gamma_{\rm rel}\simless 1$  (i.e. before \np decoupling).
This results in some differences in the dynamics close 
to the decoupling radius. We
find that the two fluids decouple over a narrower radial range (i.e. 
sharper decoupling). Furthermore, the fitting formula (\ref{sigmatot}) results in terminal
neutron Lorentz factors that are $\sim$$10\%$  higher than those found 
when we use the Rossi et al. (2006) expressions for the \np scattering cross section.

With eqs. (\ref{x}), (\ref{gammap}) and (\ref{gamman}) one has the complete
description of the dynamics of the fireball in the Thomson thick part of the
flow (i.e. below the photosphere). In the optically thin part radiation and
matter decouple and expression (\ref{adiabatic}) is no longer applicable.

Since radiation pressure is the driving mechanism of acceleration in the
fireball, one would expect no further acceleration of the flow to take place 
above the photosphere. On the other hand, although most of the photons do not 
scatter with electrons above the
photosphere, the electrons (outnumbered by the photons by a factor $\sim
10^{5}$) are still repeatedly scattered resulting in a residual acceleration
of the flow in the optically thin region. This residual acceleration is given
by the expression (Beloborodov 2002; Rossi et al. 2006, appropriately 
modified to include the neutron fluid):
\be
\frac{\rm d \Gamma_{\rm p}}{\rm d r}+\xi \frac{\rm d \Gamma_{\rm n}}{\rm d r}=
\frac{\sigma_{\rm T} L_{\rm r}}{16\pi \Gamma_{\rm p}^2
  r^2mc^3}\Big[1-\Big(\frac{\Gamma_{\rm p}r_{\rm ph}}{\Gamma_{\rm p}(r_{\rm
	ph})r}\Big)^4\Big]+\frac{\Gamma_{\rm p}-\Gamma_{\rm
    n}}{1+\xi}\frac{{\rm d}\xi }{\rm{d} r} \, ,
\label{gammapthin}
\ee
where $\sigma_{\rm T}$ is the Thomson cross section and $L_{\rm r}=16\pi
r^2c\Gamma_{\rm p}^2 e/3$ stands for the radiative luminosity of the flow.
The first term in the right hand side of the last expression accounts for the
residual acceleration from radiation and the second for the  effect of neutron
decay on the dynamics.  
Using  eqs. (\ref{LumFB}) and (\ref{eta}), we have for the radiative
  luminosity of the flow:
\be
L_{\rm r}=L\Big(1-\frac{\Gamma_{\rm p}+\xi \Gamma_{\rm n}}{\eta(1+\xi)}\Big) \, .
\label{Lrad}
\ee
The expressions  (\ref{ksir}), (\ref{gamman}), (\ref{gammapthin}) and (\ref{Lrad})
describe the dynamics of the flow in the Thomson thin regime.

\subsection{The reconnection model}

In the magnetic reconnection model the flow is considered starting from the
Alfv\'en point $r_{\rm A}$ and is dominated by Poynting flux. 
The luminosity  of the flow is the sum of the kinetic and Poynting flux:
\be
L=4\pi r^2c\Big[\Gamma_{\rm p}^2(4e/3+n'_{\rm p}mc^2)+\Gamma_{\rm n}^2n'_{\rm
    n}mc^2\Big]+c(rB)^2 \, ,
\label{Lac}
\ee 
where $B$ is the magnetic field strength in the central engine frame, 
which is dominated by its toroidal component.
 
A detailed investigation of the properties of a neutron-free flow is presented in 
Drenkhahn (2002)  under the assumption of a cold flow (i.e. a flow where the term $4e/3$
is neglected with respect the other terms in  eq. \eqref{Lac}). A full
numerical investigation showed that the dynamical description under the
cold flow assumption is rather accurate (Drenkhahn \& Spruit 2002). Hereafter, we 
assume that the flow is cold. One should keep in mind, however, that 
though of moderate dynamical significance, the internal energy of the
flow -- dominated by the energy density of radiation -- plays a crucial
role for its photospheric emission (Giannios 2006; Giannios \& Spruit 2007).    
Furthermore, the cold flow assumption can overestimate the acceleration of
the flow in the Thomson thin region by up to 50\% in the limit that the
internally dissipated energy  does not stay in the flow but is efficiently 
radiated away (Drenkhahn \& Spruit 2002). More realistically only a
fraction of the dissipated energy is radiated away and the error we make in
the Thomson thin region is smaller.

Setting $e=0$ and using  eqs. (\ref{Mdot}) and (\ref{Lac}) we have
\be
L=\frac{\Gamma_{\rm p}+\xi \Gamma_{\rm n}}{1+\xi} \dot{M}c^2+c(rB)^2=\frac{\Gamma_{\rm p}+\xi
  \Gamma_{\rm n}}{1+\xi} \dot{M}c^2 (1+\sigma) \, ,
\label{mdotL}
\ee
where $\sigma\equiv (1+\xi)(rB)^2/(\Gamma_{\rm p}+\xi \Gamma_{\rm n})\dot Mc$ is the magnetization parameter of
the flow and stands for the Poynting-to-kinetic flux ratio. 
Using the last expression, the baryon loading of the flow is
\be
\eta\equiv \frac{L}{\dot Mc^2}=\frac{\Gamma_{\rm p}+\xi
  \Gamma_{\rm n}}{1+\xi} (1+\sigma) \, .
\label{etaAC}
\ee
In the reconnection model, the flow is considered starting from the Alfv\'en
radius with magnetization $\sigma_0$. In the inner part of the flow
the very frequent \np collisions lead to $\Gamma_{\rm p,0}\simeq
\Gamma_{\rm n, 0}=\sqrt{\sigma_0}$. In view of  eq. \eqref{etaAC}, one has that
$\eta=\sqrt{\sigma_0}(1+\sigma_0)\simeq \sigma_0^{3/2}$. The initial
magnetization $\sigma_0$ is, thus, an alternative means in parameterizing the
baryon loading of a PFD flow.

The radial dependence of the magnetic field strength is given by the induction
equation that is appropriately modified to take into account the
magnetic field dissipation through reconnection (Drenkhahn \& Spruit 2002):
\be
\frac{{\rm d} (rB)}{{\rm d} r}=-\frac{rB}{c\tau_{\rm d}} \, .
\label{induction}
\ee
Here, 
\be
\tau_{\rm d}=\frac{2\pi\Gamma_{\rm p}^2}{\varepsilon
\Omega}\sqrt{\frac{\sigma+1}{\sigma}}
\ee
is the dissipation timescale 
of the magnetic field (in the central engine frame), $\Omega$ stands
for the angular frequency of the rotator, and $\varepsilon$ parameterizes the
magnetic reconnection speed $v_{\rm rec}$.   
As in most models of magnetic reconnection, $v_{\rm rec}$ scales with the
Alfv\'en speed $v_{\rm A}$, i.e. $v_{\rm rec}=\varepsilon v_{\rm A}$ 
  (see, for example, Lyubarsky 2005). A nominal value used for $\varepsilon$ is 0.1.  

By combining eqs. (\ref{ksir}), (\ref{Lac}), (\ref{mdotL}) and (\ref{induction})
one can eliminate the magnetic field $B$ and derive an equation for the 
bulk Lorentz factor of the protons and the neutrons:
\be
\frac{\rm d \Gamma_{\rm p}}{\rm d r}+\xi\frac{\rm d \Gamma_{\rm n}}{\rm d r}=
\frac{2}{c\tau_{\rm d}}\big((1+\xi)\sigma_0^{3/2}-\Gamma_{\rm p}-\xi \Gamma_{\rm
  n}\big)+\frac{\Gamma_{\rm p}-\Gamma_{\rm
    n}}{1+\xi}\frac{{\rm d}\xi }{\rm{d} r} \, .
    \label{gammapAC}
\ee
The last expression, in combination with eqs. (\ref{ksir}) and (\ref{gamman}), describes the
the dynamics of neutron-rich flows in the reconnection model.

\subsection{Results}
\label{sec:dyn:results}
 Having derived a closed system of equations that describe the dynamics of
neutron-rich flows, we proceed with the investigation of the dependence of
their properties on the parameters of the flow for both fireballs and 
strongly magnetized flows.

\subsubsection{The fireball}  
\label{sec:dyn:results:FB}
By numerically solving  eqs. (\ref{x}), (\ref{gammap}) and (\ref{gamman})
in the Thomson thick part of the flow and eqs.  (\ref{ksir}), (\ref{gamman}),
(\ref{gammapthin}) and (\ref{Lrad}) above the photosphere,
one can follow the various stages of the neutron-rich fireball
(see also Rossi et al. 2006).

\begin{figure}
\begin{center}
\includegraphics[height=8cm, angle=270]{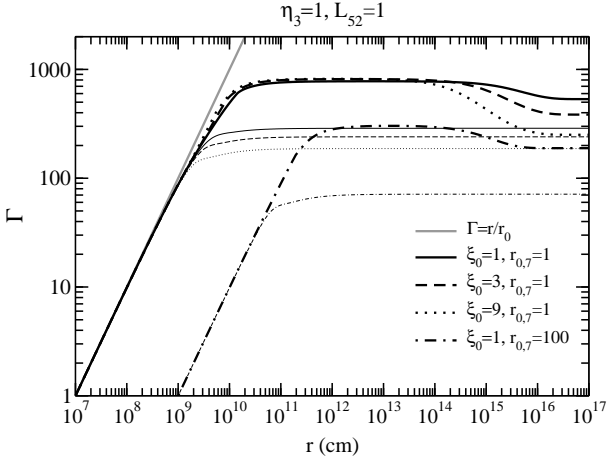}
\caption{Bulk Lorentz factor of the protons (thick lines) and neutrons
  (thin lines) for different 
values of the initial neutron-to-proton ratio $\xi_0$ and radius $r_0$ of the fireball.
At small radii, both protons and neutrons are in the linear acceleration
regime (gray line). After \np decoupling the neutrons saturate while protons
can be further accelerated by radiation pressure. At $r\sim 10^{15}$ cm the
neutrons decay into protons that interact and decelerate the preexisting protons.
\label{FB1}}
\end{center}
\end{figure}

\begin{figure}
\begin{center}
\includegraphics[height=8cm, angle=270]{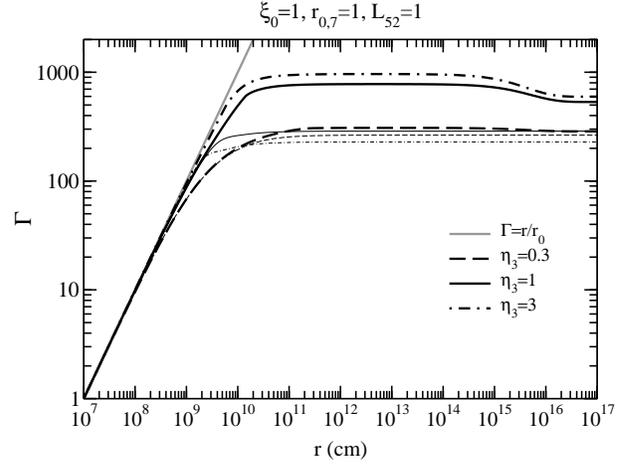}
\caption{Bulk Lorentz factor of the protons  (thick lines) and neutrons
 (thin lines) for different 
values of the baryon loading $\eta$ of the fireball. For low baryon loading
(high $\eta$), the protons are accelerated to much higher bulk Lorentz factors
  than the neutrons. For high $\eta$, the saturation of the protons takes place  
close to the Thomson photosphere while the photospheric emission is very powerful.
\label{FB2}}
\end{center}
\end{figure}

In Figs.~(\ref{FB1}) and (\ref{FB2}), the bulk Lorentz factors of the
proton and the neutron fluids are plotted as function of radius for 
different values of the parameters of the fireball model. The latter are 
the luminosity of the flow $L$, the baryon loading $\eta$, the initial
neutron-to-proton ratio $\xi_0$ and initial radius $r_0$ of the flow. 
All the models studied have $\eta\simmore 100$ relevant for GRB flows.

These low-baryon flows pass through an initial phase of rapid acceleration.  
During this phase, the neutron and proton fluids are strongly coupled
and move practically with the same bulk Lorentz factor. Setting
$\Gamma_{\rm p}\simeq \Gamma_{\rm n}$ in eq. (\ref{gammap}) we have
\be
\frac{\rm d \Gamma_{\rm p}}{\rm d r}=\frac{\Gamma_{\rm
    p}}{r}\frac{2x}{2x+3(1+\xi_0)} \, .
\ee
In the limit of $x\gg 3(1+\xi_0)/2$, radiation
pressure leads to the well known  linear acceleration of the flow as function of radius
(cf. Goodman 1986; Paczy\'nski 1986; Piran et al. 1993):
\be
\Gamma_{\rm p}\simeq \Gamma_{\rm n}=\frac{r}{r_0} \, ,
\label{gammafireball}
\ee
If no \np decoupling were to take place, the bulk Lorentz factor
of the flow would saturate at $\Gamma_{\infty}=\eta$ at the saturation radius
$r_{\rm s}=\eta r_0$. 

Note that although at small radii the numerical results follow
the linear scaling (\ref{gammafireball})  closely, there 
are deviations from this scaling appearing for $\Gamma_{\rm p}\simmore
100$ for the models presented in Figs.~1 and 2. Since
eq.~(\ref{gammafireball}) is exact for a fireball with a negligible number 
of baryons,  finite-$\eta$ flows have bulk Lorentz factors $\Gamma(r)<r/r_0$.

At larger radii the density of the flow drops and \np scatterings become
less frequent. When the comoving dynamical timescale becomes shorter than the
\np scattering timescale, the two fluids decouple and the neutrons are
not accelerated any more. The relative velocity increases  rapidly at 
decoupling. One can define the decoupling condition as $\Gamma_{\rm
    rel}\beta_{\rm rel}=1$.
Setting this condition in (\ref{gamman}) and using also (\ref{gammafireball}) 
one finds for the decoupling radius
\be
r_{\rm np}=2.6 \times 10^{9}L_{52}^{1/3}r_{0,7}^{2/3}\eta_3^{-1/3}
\Big(\frac{1+\xi_0}{2}\Big)^{-1/3}\quad\rm cm \, ,
\label{rnpFB}
\ee
and for the Lorentz factor at decoupling
\be
\Gamma_{\rm np}=2.6 \times 10^2 L_{52}^{1/3}\eta_3^{-1/3} r_{0,7}^{-1/3}\Big(\frac{1+\xi_0}{2}\Big)^{-1/3} \, .
\label{gammanpFB}
\ee 
If the flow reaches its terminal Lorentz factor  before 
\np decoupling has taken place, both the neutron
and proton flows coast with the same speed. 

For a flow with a sufficiently high
$\eta$, i.e.
\be
\eta>\eta_{\rm cr}\equiv
360L_{52}^{1/4}r_{0,7}^{-1/4}\Big(\frac{1+\xi_0}{2}\Big)^{-1/4} \, ,
\label{etacr}
\ee  
the protons  keep being accelerated after \np decoupling has taken place
while the neutrons coast with $\Gamma_{\rm n}\sim \Gamma_{\rm np}$.
The bulk Lorentz factor at  \np decoupling $\Gamma_{\rm np}$ provides
 a good estimate of the saturation Lorentz factor of the neutrons $\Gamma_{\rm
n,s}$. To quantify this statement, we have compared the analytical
estimate for $\Gamma_{\rm np}$ with the numerical values of $\Gamma_{\rm n}$
at  a large radius (here taken  at $r=10^{17}$ cm) and  found that 
the two quantities agree with each other
within $\sim$$25$\% for the (rather large) parameter space $\eta_{\rm
  cr}<\eta<3000$, $0.01<L_{52}<10$, $0<\xi_0<10$ and $1<r_{0,7}<100$.

When condition (\ref{etacr}) is satisfied, the protons 
are further accelerated  by radiation pressure after \np decoupling until either
all internal energy has been used or the flow 
crosses the photosphere, where the flow
becomes transparent with respect to Thomson scattering so that
radiation and matter decouple.

An estimate of the maximum Lorentz factor of the protons is given by 
assuming a neutron-free flow after \np decoupling with luminosity
$\hat {L}$ that does not include the kinetic energy of neutrons (i.e. $\hat 
{L}=L-\Gamma_{\rm np}\xi_0 {\dot M}c^2/(1+\xi_0 )$) and mass flux $\hat{\dot
  {M}}=\dot {M}/(1+\xi_0)$. The baryon loading of the decoupled proton flow 
 is
\be
\hat{\eta}=\frac{\hat{L}}{\hat{\dot
  {M}}c^2}=\eta(1+\xi_0)-\xi_0\Gamma_{\rm np} \, .
\ee
The acceleration of the proton fluid will saturate at
\be
\Gamma_{\rm p, s}=\rm{min}[\hat{\eta},\hat{\eta}_{\rm rad}] \, ,
\label{gammaps}
\ee 
where $\hat{\eta}_{\rm rad}=(\hat{L}\sigma_{\rm T}/4\pi r_0mc^3)^{1/4}$
gives the terminal Lorentz factor of the protons when the acceleration of the flow
is limited by  photospheric crossing (Beloborodov 2002).
This estimate takes into account the residual  acceleration in the
optically thin region discussed in section~\ref{sect:fb}.

At still larger radii of the order  of $r_{\beta}=\Gamma_{{\rm
 np}}c\tau_{\beta}\sim 10^{15}-10^{16}$ cm, neutron decay has an appreciable effect on the
dynamics of the flow. The neutrons decay into protons and interact
with the faster moving proton flow, thereby slowing it down. Note that at 
distances $10^{17}$ cm, practically
all the neutrons have decayed. The terminal Lorentz factor of the protons
there is $\Gamma_{\rm p,\infty}\le \eta$. For flows with $\hat{\eta}>\hat{\eta}_{\rm rad}$,
most of the energy is {\it not} used to accelerate the baryons (resulting in 
$\Gamma_{\rm p,\infty}\ll \eta$) but instead appears as photospheric emission
of the flow. 

Further out, the flow enters the afterglow phase
where it decelerates because of interaction with the circumburst medium. This 
last phase is not considered in this study.
  
\subsubsection{The reconnection model}
\label{sec:dyn:results:AC}

\begin{figure}
\begin{center}
\includegraphics[height=8cm, angle=270]{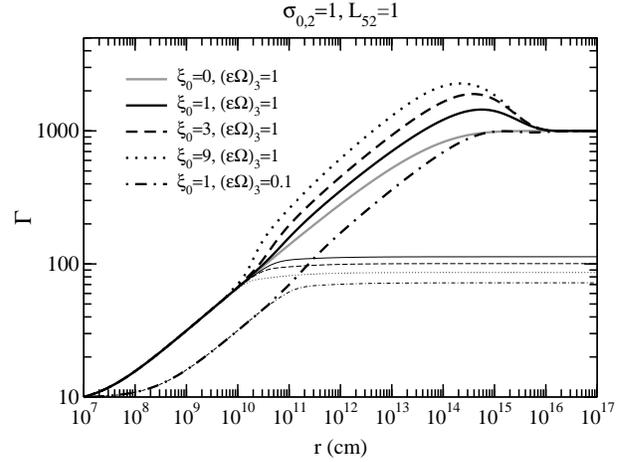}
\caption{Bulk Lorentz factors of the protons (thick lines) and neutrons
(thin lines) for different 
values of the initial neutron-to-proton ratios $\xi_0$ and reconnection speed parameterized
by $\varepsilon \Omega$ in the reconnection model.  At  \np decoupling
radius the acceleration rate of the protons is enhanced. This effect 
is particularly pronounced for $\xi_0\gg 1$ flows. At $r\sim 10^{14}-10^{15}$
cm, the neutrons decay causing deceleration of the protons.  
\label{AC1}}
\end{center}
\end{figure}

\begin{figure}
\begin{center}
\includegraphics[height=8cm, angle=270]{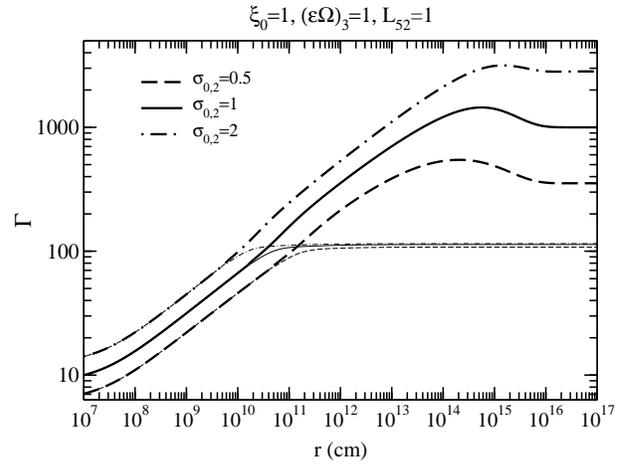}
\caption{Bulk Lorentz factors of the protons  (thick lines) and neutrons
 (thin lines) for different 
baryon loadings parameterized by the magnetization parameter $\sigma_0$ in the
reconnection model. The bulk Lorentz factor of the neutrons at \np
decoupling is essentially independent of $\sigma_0$, in agreement with the 
analytical estimate (\ref{gammanpAC}).
\label{AC2}}
\end{center}
\end{figure}

We now present the various phases of the development of the flow
in the context of the reconnection model. The neutron-free flow has been 
studied by Drenkhahn (2002) and Drenkhahn \& Spruit (2002). Here we focus on
the dynamical effect of the neutrons.
In Figs.~\ref{AC1} and \ref{AC2},  the bulk Lorentz factors of the proton
and the neutron fluids are plotted as function of radius for 
different values of the parameters of the reconnection model. These parameters are 
the luminosity of the flow $L$, the initial magnetization $\sigma_0$ of the 
flow (that also parameterizes the baryon loading since $\eta\simeq \sigma_0^{3/2}$), the initial
neutron-to-proton ratio $\xi_0$ and the combination $\varepsilon \Omega$ that parameterizes
the reconnection speed.

The flow passes through an initial phase
 of acceleration  
where the neutron and proton fluids are strongly coupled
and move practically with the same bulk Lorentz factor. Setting
$\Gamma_{\rm p}\simeq \Gamma_{\rm n}$ in 
eq. (\ref{gammapAC}) we have
\be
 \frac{\rm d \Gamma_{\rm p}}{\rm d r}=
\frac{\varepsilon \Omega\sqrt{1-\Gamma_{\rm p}/\sigma_0^{3/2}}}{\pi c\Gamma_{\rm p}^2}
\big(\sigma_0^{3/2}-\Gamma_{\rm p}\big) \, .
\label{gammaAC}
\ee
In the limit of $\Gamma_{\rm p}\ll \sigma_0^{3/2}$ (i.e. the flow is still dominated
by Poynting flux), the last equation can be integrated analytically to find (Drenkhahn 2002):
\be
\Gamma=\Big(\frac{3\varepsilon \Omega \sigma_0^{3/2}}{\pi
c}(r-r_0)+\sigma_0^{3/2}\Big)^{1/3} \, ,
\ee
The reconnection model predicts a gradual acceleration of the flow 
$\Gamma\sim r^{1/3}$ in the regime $ \sqrt{\sigma_0}\ll \Gamma_{\rm p} \ll
\sigma_0^{3/2}$ with the bulk Lorentz factor of the  flow given by
\be
\Gamma_{\rm p}\simeq \Gamma_{\rm n}=(3\varepsilon
\Omega\sigma_0^{3/2}r/\pi c)^{1/3} \, .
\label{gammapac}
\ee
This expression is valid as long as the neutrons have not decoupled from the
protons and the flow has not reached its terminal Lorentz factor
$\Gamma_{\infty}=\sigma_0^{3/2}$ at the
saturation  radius
\be
r_{\rm  s}=\frac{\pi c}{3\varepsilon\Omega}\sigma_0^3 \, .
\ee

At larger radii the density of the flow drops and nuclear scatterings become
less frequent. When the comoving dynamical timescale becomes shorter than the
\np scattering timescale, the two fluids decouple and the neutrons are
not accelerated any more. Their relative velocity $\beta_{\rm rel}$ increases rapidly around
decoupling. As for fireballs, one can define the decoupling  condition as
  $\Gamma_{\rm rel}\beta_{\rm rel}=1$. 
Setting this condition in eq. \eqref{gamman} and  using also  eq. \eqref{gammapac}
one finds for the decoupling radius
\be
r_{\rm np}=4.1\times 10^{10}L_{52}^{3/5}(\varepsilon\Omega)_3^{-2/5}\sigma_{0,2}^{-3/2}
  \Big(\frac{1+\xi_0}{2}\Big)^{-3/5} \rm\quad cm \, .
\label{rnpAC}
\ee
The bulk Lorentz factor of the flow at the decoupling is
\be
\Gamma_{\rm np}=110 L_{52}^{1/5}(\varepsilon\Omega)_3^{1/5}\Big(\frac{1+\xi_0}{2}\Big)^{-1/5} \, .
\label{gammanpAC}
\ee
If the flow reaches its terminal Lorentz factor at $r_{\rm s}$ before \np decoupling
has taken place, both the neutron
and proton flow coast with the same speed. For a flow with a sufficiently high
$\sigma_0$, such that
\be
\sigma_0>\sigma_{\rm cr}\equiv
23 L_{52}^{2/15}(\varepsilon\Omega)_3^{2/15}\Big(\frac{1+\xi_0}{2}\Big)^{-2/15} \, ,
\ee  
the protons are further accelerated after \np decoupling has taken place
while the neutrons coast with $\Gamma_{\rm n}\sim \Gamma_{\rm np}$.
The bulk Lorentz factor at \np decoupling $\Gamma_{\rm np}$ provides
a  good estimate of the saturation Lorentz factor of the neutrons
$\Gamma_{\rm n,s}$. Comparing the analytical
estimate for $\Gamma_{\rm np}$ with the numerical values of $\Gamma_{\rm n}$
at large radii (taken here at $r=10^{17}$ cm), we have found that the two quantities agree with each other
within $\sim$$10$\% for the parameter space $\sigma_{\rm
  cr}<\sigma_0<300$, $0.01<L_{52}<10$, $0<\xi_0<10$ and $0.01<(\varepsilon\Omega)_3<10$.
  
The critical value $\sigma_{\rm cr}$ corresponds to baryon loading $\eta_{\rm
cr}\simeq \sigma_{0, cr}^{3/2}\sim 100$. For baryon loadings $\eta\simmore 100$ 
relevant for GRB flows, \np decoupling takes place before
the saturation radius has been   crossed. 
In this case a substantial amount of magnetic energy is dissipated
at radii $r>r_{\rm np}$, which is used to  accelerate the 
protons.

At the \np decoupling radius the flow  becomes effectively less
baryon loaded and the protons increase their Lorentz factor more rapidly than the $\Gamma_{\rm p}\sim
r^{1/3}$ scaling. This enhanced acceleration is  particularly pronounced in neutron
dominated flows (where $\xi_0\gg 1$; see Fig.~\ref{AC1}). A similar enhancement
in the acceleration has been found by Vlahakis et al. (2003) in the
context of a different MHD model for GRBs.

Note that soon after  \np decoupling has taken place the flow crosses the Thomson photosphere.
The protons keep accelerating after the photospheric crossing in the
magnetized flow since the acceleration is magnetic and
not driven by radiation pressure as in the fireball model. At larger radii, the protons can reach bulk
Lorentz factors in excess of the limit $\sigma_0^{3/2}$ that characterizes a
pure proton flow (shown with dotted line in Fig.~3).

At larger radii the neutrons undergo  beta decay. For 
high-$\xi_0$ flows, at radius $r\sim \Gamma_{{\rm
np}}c\tau_{\beta}/\xi_0$ the number of neutrons that have decayed is
comparable with the initial number of protons in the flow and the effect of neutron
decay on the bulk motion of the protons becomes appreciable. 

After magnetic
dissipation has ceased and most of the neutrons have decayed, all the 
available energy has been transferred to the protons. The bulk 
Lorentz factor of the protons at large radii saturates to the value 
$\Gamma_{\rm p,\infty}=\sigma_0^{3/2}$. This takes place at $r\sim 10^{16}$ 
cm. At these radii the flow is expected to enter the afterglow phase
which is not considered here.

\section{Particle creation in inelastic neutron -- proton collisions}
\label{sec:particlecreation}

In the previous section we demonstrated that for low enough baryon
loading, the neutrons decouple before the acceleration of the flow is
completed in both fireballs and PFD  flows. This leads to
neutrons and protons developing relative motions and to energetic \np collisions.
Here, we study the production of pions through inelastic \np collisions
in the relativistic outflow and the subsequent decay of pions into $\gamma$-rays and neutrinos.
We present analytical estimates for the secondary particle fluences
and energies, and compare these estimates with numerical results
based on the model discussed in section~\ref{dynamics}. 

For the analytical estimates,
we approximate the proton and neutron Lorentz factors
as follows:
\be
\label{Gammapn_simp}
\Gamma_{\rm p}  \simeq  \Big(\frac{r}{r_0}\Big)^p \, ; 
\qquad
\Gamma_{\rm n} \simeq  \rm{min} \left[\Gamma_{\rm p}, \Gamma_{\rm np} \right] \, ,
\ee
where  $\Gamma_{\rm np}$ is the Lorentz factor of the flow at decoupling,
$p$ is a model parameter that allows us to
consider the fireball model and the
reconnection model together ($p=1$ for fireballs and $p=1/3$ for
the reconnection model), and $r_0$ is 
a suitable length scale. For the fireball model $r_0$ is the
initial radius where the fireball in injected, which is a free
parameter of the model. In the reconnection model
$r_0\equiv \pi c/3\varepsilon \Omega \sigma_0^{3/2}$ is a length scale 
defined by the specific combination of the parameters -- it has no deeper physical meaning but merely 
serves in rewriting the expression \eqref{gammapac} in a more compact form.

Using the unifying notation (\ref{Gammapn_simp}) for the bulk Lorentz
 factor of the protons and the neutrons, we express the \np decoupling radius and the Lorentz
factor at decoupling as:
\be
r_{\rm np}   =   \Big(\frac{\bar{\sigma} Lr_0^{2p}}{8\pi p m c^3 (1+\xi) \eta}\Big)^{\frac{1}{2p+1}} \, ;
\label{rnp}
\ee
\be
\Gamma_{\rm np}   =   \Big(\frac{\bar{\sigma} L}{8\pi r_0p m c^3 (1+\xi)  \eta}\Big)^{\frac{p}{2p+1}} \, ,
\label{gammanp}
\ee
which combines eqs. \eqref{rnpFB}, \eqref{gammanpFB}, \eqref{rnpAC} and \eqref{gammanpAC}.

\subsection{The pion production radius}
\label{sec:pionradius}
For sufficiently low baryon loading in the flow, 
pion creation in inelastic \np collisions is possible after
\np decoupling and the subsequent acceleration of the protons
with respect to the neutrons.
We define the pion creation radius $r_{\pi}$ as the minimum radius where
the relative velocity between decoupled
neutrons and protons is large enough to
create pions through inelastic \np collisions.

 The production of a secondary particle with mass $\mu$ requires center-of-mass
energy $\sqrt{s} > 2 m c^2 + \mu c^2$. 
Assuming that $\Gamma_{\rm p} (r) \gg 1$ and $\Gamma_{\rm n} (r) \gg 1$ 
at radii $r > r_{np}$, and taking the \np
collision angle  equal to zero
(tail-on collisions), we express the
center-of-mass energy $\sqrt{s}$ as
\be
\label{eq:sqrts_chi}
\sqrt{s} = m c^2 \lb \chi^{1/2} + \chi^{-1/2}  \rb \, ,
\ee
where we introduce the useful quantity
\be
\chi(r) \equiv  \frac{\Gamma_{\rm p} (r)}{ \Gamma_{\rm n} (r)} \, .
\ee
From eq. \eqref{eq:sqrts_chi}, we find that
pions (which are the lightest mesons) can only be created if
$\chi (r) > \chi_\pi $, where $\chi_{\pi^0} = 2.13$ corresponds to neutral pion production and
$\chi_{\pi^{\pm}} = 2.16$ to charged pion production. We will use the
average value $\chi_{\pi} = 2.15$ in this work.
Using the approximate proton and neutron Lorentz factors expressed in  eqs. \eqref{Gammapn_simp}, we find that
\be
r_{\pi} \simeq \chi_\pi^{1/p} \, r_{np}  \, ,
\label{rpi}
\ee
where the decoupling radius $r_{\rm np}$ is given in eq. \eqref{rnp}.
The radius from which pions can be created is thus substantially larger than the decoupling radius.
 Since the density of the flow and hence the number of \np scatterings
decrease rather steeply with radius, it is important to discriminate between
$r_{\rm np}$ and $r_{\pi}$ when considering particle creation in inelastic \np scatterings.

Pion creation by  \np interactions occurs only when the 
pion creation radius $r_\pi$ is reached before the flow saturates.
For the fireball model, saturation of {the bulk Lorentz factor of} the flow occurs
either when there is no more energy available to further accelerate the baryons
or when the flow crosses the photosphere (cf. eq. \eqref{gammaps}).
It can be shown that saturation occurs beyond the pion creation radius only
if the baryon loading of the flow is sufficiently small.
We express this condition as $\eta > \eta_\pi$, where $\eta_\pi$ is the
critical value for inelastic \np collisions to occur in the flow.
Approximating the proton and neutron Lorentz factors with
eqs. \eqref{Gammapn_simp}, we estimate that
\be
\label{eq:etapi}
\eta_\pi 
% = \lb \frac{L \bar{\sigma}}{8 \pi m c^3 r_0} \rb^{1/4}
%\frac{(\chi_\pi + \xi)^{3/4}}{1+\xi_0} 
\simeq \EE{5.1}{2}  L_{52}^{1/4} r_{0,7}^{-1/4} \psi (\xi_0) \, ,
\ee
where 
\be
\psi (\xi_0) \equiv 0.85 \, (\chi_\pi + \xi_0)^{3/4} (1+\xi_0)^{-1}
\ee
is a slowly-varying function 
normalized so that $\psi(1) = 1$. In deriving eq. \eqref{eq:etapi} we take neutron decoupling into account by
using the neutron-free luminosity $\hat{L}$ and mass flux $\hat{\dot{M}}$
as defined in section~\ref{sec:dyn:results:FB}. 
The numerical investigation of the fireball dynamics
(see section~\ref{sec:dyn:results:FB}) shows 
that the proton Lorentz factors are 
substantially below the $\Gamma_{\rm p} \propto r$
scaling solution around \np decoupling. As a result, the
proton -- neutron relative velocity is smaller and the
pion production radius is pushed outward with respect to the analytical
estimate (\ref{rpi}).
This effect makes it more difficult to create pions
in the flow and requires $\eta$ to be higher than the estimate \eqref{eq:etapi}.
Using the numerical model discussed in section~\ref{dynamics}
we find that inelastic \np collisions in fireballs
occur generally when $\eta /\eta_\pi\gtrsim 2 $, where $\eta_\pi$ is expressed in eq. \eqref{eq:etapi}.
For neutron-rich flows ($\xi_0 \gtrsim 3$), neutron decoupling results in a relatively pure
flow so that the protons follow the scaling approximation \eqref{eq:etapi} more
closely and
inelastic \np collisions occur already when $\eta/\eta_\pi \gtrsim 1.5 $.
Nevertheless, these results place quite stringent conditions on the fireball
model parameters so that only a small fraction of GRB fireballs is expected to exhibit
inelastic \np collisions between bulk protons and neutrons.

For the AC model we find that, similar to the fireball case,  inelastic
\np collisions only occur for a sufficiently low
baryon loading. We express this as $\sigma_0 > \sigma_{0,\pi}$, where we 
use eqs. \eqref{Gammapn_simp} to estimate that
\be
\label{eq:sigmapi}
\sigma_{0,\pi}
%\equiv
%\chi_\pi^{2/3} \lb \frac{9 \bar{\sigma} L \ep \Omega  }{8 \pi^2 m c^4 ( 1+ \xi_0)} \rb^{2/15}
\simeq 38 \times L_{52}^{2/15} \lb \ep \Omega \rb_3^{2/15} 
\lb \frac{2}{1+\xi_0}\rb^{2/15} \, .
\ee
We find that eq. \eqref{eq:sigmapi} is consistent with the critical
value for $\sigma_0$ obtained from numerical results on the 
proton and neutron dynamics (using the numerical model described in  section~\ref{dynamics}).
This value of $\sigma_{0,\pi}$ corresponds to a critical baryon loading
for inelastic \np collisions $\eta_{\pi}=\sigma_{0,\pi}^{3/2}\sim 230$ which
is much lower than the critical value required in fireballs. Inelastic \np collisions
thus take place for a larger range of the parameter space in the reconnection
model with respect to the fireball. 

The strength of any neutrino and $\gamma$-ray emission that is a result
of the decay of the products (mainly pions) of these collisions depends
critically on the optical depth to inelastic \np scattering. The
calculation of this optical depth is the topic of the next section.

\subsection{Optical depth for inelastic \np collisions}
\label{sec:tau}
The optical depth ${\rm d}\tau$ for a neutron with velocity $c \beta_{\rm n}$ to
scatter inelastically with
a population of protons with velocity $c \beta_{\rm p}$ and proper density $n'_{\rm p}$
within $r \ldots r + {\rm d}r$ is given by 
(see, e.g., Landau and Lifshitz 1971)
\be
\label{eq:dtaudr}
{\rm d} \tau = \sigma_{\tinel}  \Gamma_{\rm p} n'_{\rm p}  \lb \frac{\beta_{\rm p} - \beta_{\rm n}}{\beta_{\rm n}} \rb {\rm d} r
\simeq  \frac{\sigma_{\tinel}  n'_{\rm p} }{2 \Gamma_{\rm n}} \lb \chi - \frac{1}{\chi} \rb  {\rm d} r \, ,
\ee
where we assume in the last approximation that $\Gamma_{\rm p} \gg 1$ and  $\Gamma_{\rm n} \gg 1$ and
that the collisions are tail-on.

At low center-of-mass energies the elastic and inelastic \np cross sections are energy dependent. 
We find that for $\chi_\pi \leq \chi \lesssim 10$
(which is the range of interest here)
the elastic cross section is well described with
$\sigma_{\tel} (\chi) = 0.75 \bar{\sigma}/ \ln \chi$, where
$\bar{\sigma} \equiv \EE{4}{-26} $ cm$^2$. A comparison between this
approximation and experimental data on the elastic cross section
taken from Yao et al. (2006)
is presented in appendix \ref{sec:app:sigma}. 
In the following, we 
express the inelastic \np cross section as
\be
\label{eq:sigmainel}
\sigma_{\tinel} (\chi > \chi_\pi) =  \bar{\sigma} \lb 1 - \frac{0.75}{\ln \chi }\rb \, .
\ee

We note here that the energy dependence of the \np inelastic cross section
has an important effect on the optical depth. If one assumes a
constant cross section $\sigma_{\tinel} = \EE{3}{-26}$
cm$^2$ (as is often done in the literature) the optical depths are larger by a factor $\sim$4 for both the fireball
model and the reconnection model. Hence, the more
realistic cross section adopted in this work  leads to substantially
lower estimates for the number of created particles.

We consider, in general, the situation that neutrons coast with a constant Lorentz
factor $\Gamma_{\rm n}$ while protons are accelerated up to infinity with a Lorentz factor
$\Gamma_{\rm p} \propto r^p$. Keeping $p$
as a free parameter, we integrate eq. \eqref{eq:dtaudr} through the flow to find that
\be
\label{eq:tau}
\tau(p) = \int_{\chi_\pi}^{\infty} {\rm d} \chi \lb 1 - \frac{0.75}{\ln \chi} \rb
\lb \chi^{-1} - 
\chi^{-3} \rb \chi^{-1/p} \, ,
\ee
where eq. \eqref{rnp} was used to eliminate all parameters but $p$.
We thus find that the optical depth  for inelastic \np scattering is
independent of any model parameters but the dynamical power-law index $p$.
This result is valid for all outflows with
$\Gamma_{\rm p} \propto r^p$ and $\Gamma_{\rm n} =$ const, provided that
$r_{\rm s} \gg r_\pi$. When $r_{\rm s} \gtrsim r_{\pi}$, such as in the fireball
model, it represents an upper limit.

For fireballs ($p=1$) we find from eq. \eqref{eq:tau} that
$\tau^\tFB < 0.2$, which is an upper limit because the flow saturates close to the decoupling radius.
The situation is complicated by the fact that saturation of the fireball can be due to energy requirements
or due to crossing of the photosphere.  The numerical results presented in
section~\ref{sec:dyn:results:FB} indicate that both effects cause the flow to
accelerate considerably more slowly than the scaling approximation $\Gamma_{\rm p}
\propto r$ near the pion creation radius $r_{\pi}$. This
pushes the pion creation radius outward and decreases the optical depth   for inelastic \np scattering.
We compute the optical depth numerically by a straightforward numerical integration of eq. \eqref{eq:dtaudr}
using the values of $\Gamma_{\rm p} (r)$ and $\Gamma_{\rm n} (r)$ obtained
with the numerical model\footnote{ In the numerical analysis, we use
a more accurate but also more elaborate approximation (see appendix  \ref{sec:app:sigma}) for the cross section than
the one given in eq. \eqref{eq:sigmainel}, which results in lower optical depths.
Because dynamical effects, discussed in the text, have a larger influence on the optical depth
we use expression \eqref{eq:sigmainel} for simplicity to derive an analytical
estimate.} discussed in section~\ref{dynamics}.
We find that for the parameter space
$3.5 \eta_{\pi} < \eta < 5000$, $0.01 < L_{52} < 10$, $0.3 < \xi < 10$, and 
$1 < r_{0,7} < 100$ the optical depth is approximated to within $\sim$25\% by
\be
\label{eq:numtauFB}
\tau^{\tFB} \simeq 0.11  \lb 1 - \frac{2 \eta_\pi}{\eta} \rb \, .
\ee
 In the (rather favorable for frequent inelastic scatterings) case where
  $\eta = 5000$ and $\xi=5$, the optical depth is $\tau^{\tFB} \simeq 0.1$.
For lower values of the baryon-loading parameter ($\eta / \eta_{\pi} < 3.5 $) the optical depth
is smaller than the value given in eq. \eqref{eq:numtauFB}.
A representative value for a fireball with
$\eta$ a few times the critical value  $\eta_\pi$ is $\tau^{\tFB} \simeq 0.05$.

For the reconnection model the saturation radius $r_{\rm s}$ is typically much larger than $r_\pi$. 
We can therefore estimate the optical depth $\tau^{\tAC}$ for an inelastic
\np interaction assuming that the protons are accelerated to infinity.
(In principle this overestimates the interaction probability, but the difference
is very small because the interaction probability decreases rapidly with $r$.)
Inserting $p=1/3$ in eq. \eqref{eq:tau}, we find  that $\tau^{\tAC} \simeq \EE{8}{-3}$.
This value is consistent with numerical results for flows with $\xi_0 \sim 1$.
For reference values of the parameters $L_{52}  = \xi_0 = \sigma_{0,2} = (\ep \Omega)_3  =1 $, we find also numerically that
$\tau^{\tAC} = \EE{8}{-3}$. For high values of $\xi_0$ (neutron-rich flows),
 the  extra acceleration
of the flow after neutron decoupling (discussed in section~\ref{sec:dyn:results:AC}) 
increases the optical depth by a factor few. We find that for the parameter space
$1.5 < \sigma_0 / \sigma_{0,\pi} < 10$, $0.01 < L_{52} < 100$, $0.3 < \xi_0 < 10$, and 
$0.01 < (\epsilon \Omega)_3 < 10$ the optical depth is approximated to within $\sim$25\% by
\be
\label{eq:tACnum}
\tau^{\tAC} \simeq 0.01 \, \xi_0^{1/2} \,.
\ee
 In particular, the optical depth increases to
$\tau^{\tAC} \simeq 0.03$
for very neutron-rich flows ($\xi_0 \simeq 10$). For 
$ 1 < \sigma_0 / \sigma_{0,\pi} < 1.5 $ pion creation is marginally possible and
the optical depth is smaller than the value obtained by eq. \eqref{eq:tACnum}. For
very pure flows ($\sigma_{0}/\sigma_{0,\pi} \gtrsim 10$) neutrons decouple very early
(before power-law acceleration $\Gamma \propto r^p$ is reached), which results
in an optical depth smaller by a factor $\sim$$2$ than the estimate given in eq. \eqref{eq:tACnum}.

The obtained optical depth for inelastic \np collisions is the first
 step in calculating the fluences of 
secondary pions and their decay products. The calculation of the fluences and energies
of stable decay products requires a model for the average number and average energy 
of neutrinos and $\gamma$-rays created by  \np interactions. In the
following sections we consider in detail 
the production of pions and the subsequent decay into neutrinos and $\gamma$-rays.

\subsection{Pion production}
\label{sec:pions}
For the collisions studied in this work, the typical incident energy of the proton
measured in the rest frame of the neutron is $p_{\rm p}' \sim$ 1 GeV/c. In this regime
experimental data on pion creation in \np collisions is scarce and there is no unambiguous theoretical framework.
The available data  (in particular, Prokoshin and Tiapkin 1957,
Kleinschmidt et al. 1980, Daum et al. 2002; see also electronic data files available at the PPDS website
\texttt{http://wwwppds.ihep.su:8001/}) show a rapid rise in the 
single-pion cross sections just above threshold, and indicate that two-pion
exclusive production cross sections are comparable to one-pion exclusive cross
sections for incident proton energies
(as observed in the neutron rest frame $K'$) $p_{\rm p}' \sim$ 2 GeV$/c$. Furthermore,
the ratio of $\pi^0: \pi^\pm$ depends on energy; the ratio
$\pi^- : \pi^+$ is $1 : 1$ under the assumption of nuclear isospin symmetry.

Based on the available data for incident proton energies
$p_{\rm p}' \sim 1$ GeV$/c$ we model the energy distribution and
average number of pions resulting from a
\np collisions as follows. We estimate the ratio of created pions as $\pi^0:
\pi^+ : \pi^- = 2:1:1$.
  Hence the average number of neutrinos\footnote{Here and in the
  following  $\nu_\mu$ denotes both muon-neutrinos and -antineutrinos (and
  similar for electron-neutrinos).}
and $\gamma$-rays resulting from a single inelastic \np collision
is:
\be
\label{eqs:Ngnu}
\mathcal{N}_\gamma  =  1.0 \, ; \quad
\mathcal{N}_{\nu_\mu}  =  1.0  \, ; \quad
\mathcal{N}_{\nu_e}  =  0.5  \, .
\ee
Experimental data indicate that,
for incident proton energies $p_{\rm p}' = 1.14$ GeV$/c$, 
the distribution of kinetic energy $T \equiv E - m_\pi c^2$
for $\pi^+$ mesons
peaks around 0.6 $T_{\tmax}$, where $T_{\tmax}$ is the maximum
kinetic energy that can be carried by the pion (Kleinschmidt et al. 1980).
For $\pi^0$ mesons  in the center-of-mass (COM) frame this ratio is almost unity
below $p_{\rm p}' < 1.06$ GeV$/c$ and decreases to 
$T_{\mathrm{peak}} = 0.5 \, T_{\tmax}$ at $p_{\rm p}' = 1.29$ GeV$/c$
(Prokoshin and Tiapkin 1957). Here we
assume a constant fraction of 0.6 for all pion species
and take the average pion energy for a single \np collision
in the COM frame $K''$ equal to the
peak energy: 
\be
\label{eq:pionenergy1}
\left< \ep''_{\pi} \right> =
\ep''_{\pi,\mathrm{ peak}} =  0.6 \ep''_{\pi,\mathrm{max}} + 0.4 m_\pi c^2 \, ,
\ee
where the maximum pion energy is equal to
\be
\label{eq:pionenergy2}
\ep''_{\pi,\mathrm{max}} = \frac{s - 4 m^2 c^4 + m_\pi^2 c^4}{2 \sqrt{s}} \, ,
\ee
and eq. \eqref{eq:sqrts_chi} relates the center-of-mass energy $\sqrt{s}$ to $\chi$.
In these equations, the parameter $\chi$ provides the only reference to where the collision has occurred
in the developing flow.

When the angular distribution of pions in the COM frame is known, one can derive the
full particle distribution of the decay products and transform this to the observer frame in order to
find the secondary energy as observed on earth.  However, there is to our knowledge no accurate parameterization of the angular
distribution of secondary pions created in \np collisions. In the absence of such a parameterization we
estimate the average observed energy of neutrinos by boosting to the 
observer frame from
an intermediate frame in which the secondary particles are assumed to be isotropic.\footnote{We note that, in the literature,
there are various choices regarding the frame (e.g. the neutron rest frame or
the proton rest frame) in which the energy distribution
of $\gamma$-rays and neutrinos is computed before applying the final boost to the observer frame. 
Any intermediate frame leads to the same results in the observer frame provided that the angular structure of the
particle distributions is taken into account. If an isotropic distribution is assumed, the choice of intermediate
frame is important and depends on the physics.}

Pions are created approximately isotropically in the COM frame of the \np collision. 
When neither pions nor their decay products are affected by the flow, as is the case
for neutrino production in the fireball model, the distribution
of the daughter particles can be taken to be isotropic in the COM frame. In the AC model however,
the strong magnetic field deflects the charged pions significantly
since the pion gyration period is much shorter than the pion decay time. 
We assume that in this case the pions will be distributed isotropically in the frame comoving with the proton
fluid. (Any randomized component of the magnetic field will further contribute to
isotropization in this frame).
Furthermore, in both the fireball model and the AC model $\gamma$-rays from neutral pion decay will interact with
the soft photon field of the flow, resulting in the emission of lower-energy
photons. In the following sections these issues are discussed and estimates are presented
for typical neutrino and $\gamma$-ray energies.

The decay of a charged pion also yields one electron or positron with energy $\sim$$35$ MeV. These
contribute to the $\gamma$-ray emission which is discussed in section~\ref{sect:gammarepr}.

\subsection{Observed neutrino energy}
\label{sec:obsnuenergy}
In the fireball model pions do not interact significantly with the flow so that neutrinos from charged
pion decay can be taken to be  distributed isotropically in the COM frame
$K''$  of the \np collision. For a given value of $\chi$,  the average energy of the produced neutrinos in this frame is then
\be
\label{eq:pionenergy3}
\left<\ep_\nu''\right> =  \frac{\left< \ep''_\pi \right> \ep_{\nu}^0}{m_\pi c^2}\, ,
\ee
where $\ep_{\nu}^0 \simeq 35$ MeV denotes the average neutrino energy in the rest frame of the decaying
pion. 
The average observed energy $\left< \ep_\nu \right>$ is
obtained by boosting to the observer frame with Lorentz boost factor
$\Gamma_{\mathrm{COM}} = \sqrt{\Gamma_{\rm p} \Gamma_{\rm n}}$
(appendix \ref{app:frames} contains a summary of frames and Lorentz factors
used in this work):
\be
\label{eq:pionenergyFB}
\left< \ep_{\nu}^{\rm FB} \right> =
\frac{\Gamma_{\mathrm{COM}}}{1+z}\frac{\left< \ep''_\pi \right> \ep_{\nu}^0}{m_\pi c^2} 
\simeq
\frac{\Gamma_{\rm np} \, \chi^{1/2}}{1+z}\frac{\left< \ep''_\pi \right> \ep_{\nu}^0}{m_\pi c^2} 
\, ,
\ee
where $z$ is the redshift of the source. In the last equality, we approximate the proton and neutron
dynamics by eqs. \eqref{Gammapn_simp}.

For fireballs the flow saturates close to the pion
creation radius and the bulk of the collisions occur when $\chi $ is equal to the
saturation value $\chi_{\rm s}$.
For flows with large $\eta \gtrsim 700$, where saturation is reached due to crossing of the photosphere, 
the terminal Lorentz factor of the flow equals $\Gamma_{\rm p,s} = \hat{\eta}_{\rm rad}$
(see eq. \eqref{gammaps}).
In this case, the critical value $\chi_{\rm s}$ can be estimated using the scaling law expressed in 
eqs. \eqref{Gammapn_simp}:
\be
\label{eq:chisest}
\chi_{\rm s} \simeq 4.0 \times L_{52}^{-1/12} r_{0,7}^{-1/12} \eta_3^{1/3} \lb \frac{1 + \xi}{2} \rb^{1/3} \, .
\ee
For lower values of $\eta$, the saturation value  $\chi_{\rm s}$ is smaller and reduces to the
threshold values $\chi_{\rm s} = \chi_\pi 
\equiv 2.15$ when $\eta = \eta_\pi$.
Adopting the value $\chi = \chi_{\rm s} \simeq 4$ we find from eqs. \eqref{Gammapn_simp},
\eqref{eq:pionenergy1} and \eqref{eq:pionenergyFB}
that the neutrino energy in the observer frame can be expressed as
\be
\label{eq:obsenergy}
\left< \ep_{\nu} \right> = \frac{\alpha \, \Gamma_{\rm np} \ep_{\nu}^0}{1+z} \, ,
\ee
where $\alpha^{\tFB} \simeq 4.5$ accounts for the non-zero kinetic energy of pions when they are
created  and for the fact that the particle distribution is not isotropic in the neutron
rest frame. Using the same parameter range as in section~\ref{sec:tau} we find from a numerical
analysis that $\alpha^{\tFB}$ should be slightly higher than this estimate and we will adopt
$\alpha^{\tFB} \simeq 6$ in the following.

In the AC model the situation is more complex because charged pions interact with the flow before decay
and because pions are created at various radii in the flow.
Since the pion gyration time is much  shorter than both the synchrotron
cooling time and  their lifetime, pions
will isotropize in the frame $K'$ comoving with the proton fluid without significant energy loss.
In this frame, the secondary pions are injected with energy
$\langle \ep_\pi' \rangle = \Gamma_{\rm p}'' \langle \ep''_{\pi} \rangle$, where $\Gamma_{\rm p}'' = \sqrt{s} / (2 m c^2)$
is the Lorentz factor of the incident proton as observed in the COM frame.
The observed neutrino energy is then given by the following expression:
\be
\label{eq:pionenergyAC}
\left< \ep_{\nu}^{\rm AC} \right> =
\frac{\Gamma_{\rm p}  \Gamma''_{\rm p}}{1+z}\frac{\left< \ep''_\pi \right> \ep_{\nu}^0}{m_\pi c^2} 
\simeq
\frac{\Gamma_{\rm np} \, \lb \chi^{1/2} + \chi^{3/2} \rb}{2 (1+z)}\frac{\left< \ep''_\pi \right> \ep_{\nu}^0}{m_\pi c^2} 
\, ,
\ee
where we approximate the proton and neutron dynamics by eqs. \eqref{Gammapn_simp} in the last equality.
Note that the interaction with the flow results in a substantial increase in the observed energy of the
secondary particles.
 
For flows described by the AC model \np collisions occur at various radii with different collision
energies and different values for the Lorentz boost factor $\Gamma$.
Therefore, we should average the observed energy given in
eq.  \eqref{eq:pionenergyAC} over the developing outflow. 
We express the probability for an interaction to occur while
$\chi$ is in the range $\chi \ldots \chi + {\rm d} \chi $ as $\tau(\chi) {\rm d} \chi$.
Since the scaling approximations \eqref{Gammapn_simp}
describe the flow around decoupling quite well in the AC model,
we use equation \eqref{eq:tau} to estimate that
\be
\label{eq:dtaudchi}
\tau(\chi) \equiv \frac{{\rm d} \tau}{{\rm d} \chi}  =  \lb 1 - \frac{0.75}{\ln \chi} \rb
\lb \chi^{-4} - \chi^{-6} \rb \, .
\ee
Averaging eq. \eqref{eq:pionenergyAC} over this distribution
we find that the observed neutrino energy can be expressed as in eq. \eqref{eq:obsenergy} with
$\alpha^{\tAC} \simeq 20$. This is in good agreement
with numerical results in the same parameter range as in section~\ref{sec:tau}.

\subsection{Reprocessing of $\gamma$-rays: pair cascades versus synchrotron cooling}
\label{sect:gammarepr}

While the flow is optically thin with respect to the emitted neutrinos resulting from 
charged pion decay, this is not the case for the $\gamma$-ray photons that are
produced by neutral pion decay. In the proton rest frame, the $\gamma$-rays are injected with 
average energy
(for a given value of $\chi$)
\be
\langle \ep_\gamma' \rangle =\frac{\Gamma''_{\rm p} \left< \ep''_\pi \right> \ep_\gamma^0 }{m_\pi c^2} \, , 
\ee
where $\ep_\gamma^0 =  70$ MeV. Integrating over the developing flow as in the previous section, 
we express $\langle \ep_\gamma' \rangle \simeq \beta  \ep_\gamma^0$ and estimate analytically that
$\beta \simeq 3$ for both the fireball and the AC model. This is consistent with numerical results.
Hence $\gamma$-rays have a typical energy
$\simeq 70\beta\sim 200$ MeV in the proton rest frame and are
ejected at radii $r\simmore r_{\pi}$; not far from the Thomson
photosphere of the flow. 

At these radii both fireballs and reconnection
flows carry a soft photon field with characteristic comoving energy   
in the $\sim$$1$ keV range (see Derishev et al. 1999a and Giannios 2006 
for the fireball and reconnection model, respectively). Because of this intense  
soft photon field the flow is very optically thick with respect to these $\sim$$200$ MeV
photons,  which are scattered and create pairs. In addition to 
the pairs that come from neutral pion decay,  one energetic electron (or
positron) is injected in the flow for every charged pion decay. We have included 
this contribution in the calculations that follow.

In the fireball model the dominant cooling mechanism of the electron-positron
pair is inverse Compton scattering. The upscattered soft 
photons create more pairs resulting in pair cascades. In the reconnection
model the flow is dominated by Poynting flux and the energy density of the 
magnetic field is much higher than the radiation energy density (see also eq.~(10) in 
Giannios 2006). As a result, the first generation of produced pairs cool down mainly 
through synchrotron emission. We discuss the outcome  of the $\gamma$-ray 
injection separately for the two models.

\subsubsection{Pair cascades in fireball}  
\label{sec:paircascades}
Photons in the fireball are upscattered by pairs and absorbed by soft photons
during the pair cascade. In the case of a saturated cascade, where all 
upscattered photons are absorbed, about $\sim$$10$\% of the energy of the 
$\gamma$-rays can be converted into rest mass of the pairs (Svensson 1987).
More realistically the cascade is expected to be unsaturated, converting a few 
times less energy into rest mass of pairs (Derishev et al. 1999a;  Belyanin et
al. 2003). 
    
The result of these pair cascades is twofold. Each injected $\gamma$-ray photon
is reprocessed to multiple softer photons and the flow is loaded with pairs
that contribute to its opacity. Although the saturation point
of the cascade depends on the shape of the soft photon spectrum, we
roughly estimate that photons with energies $\ep' \sim 3$ MeV in the proton rest
frame are able to escape (Belyanin  et al. 2003).
The overall emitted spectrum will be broad and most energy is emitted with observer
energies in the range $\ep \sim \Gamma_{\rm p,s}
(\ep'...10 \ep')/(1+z )\sim (2...20)/(1+z)$ GeV. The strength of this component and its
detection prospects are discussed in the next section.
    
To estimate the importance of pair loading in the flow, one should compare 
the number of produced pairs with the
number of  electrons (or, equivalently, protons) pre-existing in the flow. The flow has $\xi_0$ neutrons per
proton out of which a fraction $\tau$ scatters inelastically. This results
in $\xi_0 \tau$ inelastic scatterings per proton. Every scattering
results on average in $\sim$1 $\gamma$-ray photon (see eq.~(\ref{eqs:Ngnu})) 
with a typical energy $70\beta$ MeV in the proton rest frame.
A fraction $f\sim 3$\% of this energy is used in rest mass of pairs
which results in $\sim 70\beta f$ pairs per $\gamma$-ray. 

By setting $\xi_0=1$ and using the values of $\beta$ and $\tau$ relevant for
the reference values of the parameters for a fireball 
(see  section~\ref{sec:tau} and the beginning of this section), one finds that the pair cascades result in
$\sim$$0.15$ pairs per proton. For the neutron-dominated case where $\xi_0=5$ we find
significantly more pair loading, viz. $\sim$$2$ pairs per proton. Note that we find
significantly less pair loading of the flow because of pion decay 
compared to previous works. The main source for this discrepancy comes 
from the fact that, as we have shown in
 section~\ref{sec:tau}, the optical depth for inelastic \np scattering is about one order of magnitude smaller
than the value $\sim$$1$ that is typically assumed in these studies.

Most of these pairs are produced
at  large radii -- and low densities -- so that they do not annihilate
but stay in the flow. For $\xi_0\simless 1$, the contribution to the opacity  from
pair loading is at most moderate. On the other hand, for
$\xi_0\gg 1$ the number of produced pairs
exceeds that of the pre-existing electrons. A fraction
of those are produced below the Thomson photosphere and its location is pushed
to larger radii. This can have some backreaction on the dynamics of fireballs
that can be accelerated to slightly higher bulk Lorentz factors than those
calculated in section~\ref{sec:dyn:results:FB}, where this effect is neglected.
   
\subsubsection{Synchrotron cooling in the magnetized flow}
\label{sec:synchrotron}

We now turn our attention to the reconnection model. 
The typical energy of the electron-positron pair produced by 
scattering of a $\gamma$-ray (resulting from neutral pion decay)
with a soft photon is $\sim$$120$ MeV which corresponds to  a random electron
Lorentz factor $\gamma_{\rm e}\simeq 200-300$. The produced pair finds itself 
in a strongly magnetized flow with comoving 
$B'\simeq \sqrt{L/c r^2\Gamma_{\rm p}^2}\sim 10^6$ G for typical values of the parameters
and for the radii where most of the pion creation takes place.   

Under these conditions, the synchrotron cooling timescale of the pair 
$t'_{\rm s}\sim 10^{-6}$ sec is much shorter than the Compton 
cooling timescale. The lack of pair cascades leads to negligible pair
loading of the flow. The peak of the synchrotron emission  is located  at
$\epsilon'_{\rm s}= e\hbar B'\gamma_{\rm e}^2/m_e c\sim$$0.2...2$  keV in the 
proton rest frame. At the radii where most of the pion production 
takes place, the bulk Lorentz factor of the protons is $\Gamma_{\rm p}\sim 400-500$
which results in observer synchrotron peak in the sub-MeV energy range.
Keeping the rest of the parameters fixed to their reference
values, we find that the synchrotron emission peaks at observer energy $\epsilon_{\rm s} \simeq 120$ keV for $\xi_0=1$ and
at  $\epsilon_{\rm s}\simeq 600$ keV for $\xi_0=5$.  The spectrum is  characteristic of fast (synchrotron) cooling particles
with an exponential  cutoff above the peak and a low-energy spectral
slope of $-1/2$. The strength of this component and its detection prospects are given
in the next section.

\section{Detection prospects}
\label{sec:detection}
Using the results obtained in the previous section on the number and energy of secondary neutrinos
and $\gamma$-rays created in inelastic \np interactions, we discuss the detection prospects here.

\subsection{Neutrinos}
We express the neutrino fluence as observed on earth as
\be
\label{eq:det:flux}
\Phi_{\nu} = \frac{N_{\rm n} \mathcal{N}_{\nu} \mathcal{P}_{\rm np}}{4 \pi D_p^2} \, ,
\ee
where $\mathcal{N}_{\nu}$ 
is the average number of neutrinos created per inelastic  \np
interaction (we add the contribution of muon- and electron-(anti)neutrinos given in eq. \eqref{eqs:Ngnu} here),
$\mathcal{P}_{\rm np}$ is the inelastic \np interaction probability,
$D_p$ is the proper distance, and
\be
N_{\rm n} = \frac{\xi_0}{1+ \xi_0} \frac{E}{\eta m c^2} = \EE{3.3}{52} \lb
\frac{2 \xi_0}{1+\xi_0} \rb  E_{53} \eta_3^{-1}
\ee
denotes the number of neutrons contained in the outflow.
In the last equation, $E$ denotes the total isotropic equivalent energy of
the burst. Since $\mathcal{P}_{\rm np} \ll 1$ we express
$\mathcal{P}_{\rm np} \simeq \tau$, where $\tau$ denotes
the optical depth for inelastic \np collisions.

We consider the optimistic case of a nearby energetic burst at redshift $z=0.1$.
Assuming a universe that consists of matter and a cosmological constant, the proper distance $D_p$
is given by the following expression:
\beqs
\label{eq:DPz}
D_p = \frac{c}{H_0} \int_0^z \frac{d z'}{\sqrt{\Omega_{\Lambda,0} + \Omega_{m,0} (1+z')^3}} \, ,
\eeqs
where $\Omega_{\Lambda,0}$ and $\Omega_{m,0}$ denotes the current density parameters of
the cosmological constant and matter, respectively, and $H_0$ is the Hubble parameter.
Using the currently favored values $\Omega_{\Lambda,0} = 0.76$,  $\Omega_{m,0} = 0.24$,
and $H_0 = 73$ km s$^{-1}$ Mpc$^{-1}$ (Yao et al. 2006) we find a proper distance
$D_p = \EE{1.2}{27}$ cm. Inserting this in eq. \eqref{eq:det:flux}
we find the following neutrino particle fluences for the two models:
\be
\Phi_{\nu}^{\tFB} \simeq  10^{-4} \lb \frac{\tau}{0.05} \rb  \lb \frac{2 \xi_0}{1+\xi_0} \rb E_{53} \eta_3^{-1}  \textrm{ cm}^{-2} \, ; 
\ee
\be
\Phi_{\nu}^{\tAC} \simeq \EE{2}{-5}  \lb \frac{\tau}{0.01} \rb  \lb \frac{2 \xi_0}{1+\xi_0} \rb  E_{53} \sigma_{0,2}^{-3/2}  \textrm{ cm}^{-2} \, .
\ee
As discussed in section~\ref{sec:tau}, a typical value for the inelastic \np optical depth in the fireball model
is $\tau^{\tFB}=0.05$ and for the reconnection model $\tau^{\tAC}=0.01$. The dependence on the model parameters, as obtained
from a numerical analysis, is expressed in eqs. \eqref{eq:numtauFB} and \eqref{eq:tACnum}.

From eq. \eqref{eq:obsenergy},
the average neutrino energy  as observed on
earth is equal to
\be
\label{eq:det:enu}
\left< \ep_{\nu} \right> = \frac{\alpha \, \Gamma_{\rm np} \ep_{\nu}^0}{1+z} \, ,
\ee 
where $\alpha$ is a numerical factor that
accounts for the non-zero kinetic energy of pions when they are
created  and for the fact that the particle distribution is not isotropic in the neutron
rest frame while we boost with $\Gamma_{\rm np}$ to the observer frame.
Based on the results found in section~\ref{sec:obsnuenergy},
we take $\alpha^{\tFB} = 6$ and $\alpha^{\tAC} = 20$ for the fireball model and the AC model,
respectively. Using eqs. \eqref{gammanpFB} and \eqref{gammanpAC} for the Lorentz factors at decoupling we find
that $\langle \ep^{\tFB}_{\nu} \rangle \simeq 50$ GeV and $\langle \ep^{\tAC}_{\nu} \rangle \simeq 70$ GeV for reference values of the parameters
and a burst at redshift $z=0.1$. These values depend only mildly on the parameters through the Lorentz factor at decoupling
$\Gamma_{\rm np}$ but the value of $\alpha$ may change by a factor $\sim$$2$
depending on the burst parameters.

Following Bahcall and M\'esz\'aros (2000) we estimate the
number of interactions $R_{\nu}$ in a large-volume neutrino detector
due to the diffuse background  as
$ R_{\nu} = \Phi_{\nu} \mathcal{R}_b \sigma_{\nu} N_t $,
where $\mathcal{R}_b = 10^3 \mathcal{R}_{b,3}$  denotes the burst rate
per year, $\sigma_{\nu} = \EE{5}{-39} \, (\ep_{\nu} / \mathrm{1 \,GeV})$ cm$^2$ is the
neutrino interaction cross section and $N_t = 10^{39} N_{t,39}$ is the number of target protons in the detector.
For reference values of the parameters and an average redshift $z=1$  we find that
$ R_{\nu}^{\tFB}  \simeq 0.3$ year$^{-1}$ and $ R_{\nu}^{\tAC}  \simeq 0.07$ year$^{-1}$ for the fireball
model and the AC model, respectively. Note that, for comparison with the literature, this estimate
relies on the rather optimistic reference value of $1000$  bursts
per year leading to neutrinos through inelastic \np collisions.

The predicted diffuse neutrino detection rate for the fireball model
is a factor~$\sim$$5$ smaller than the results found by
Bahcall and M\'esz\'aros (2000). This is primarily due to the more accurate
cross sections used in this work and the distinction between \np decoupling
radius and pion creation radius. This distinction also implies that
the condition for  inelastic \np collisions to occur (as expressed in eq. \eqref{eq:etapi})
is more stringent than the condition presented by
Bahcall and M\'esz\'aros (2000). Therefore, the fraction of GRBs for which
\np decoupling occurs is expected to be much smaller and the
reference value $\mathcal{R}_b = 10^3$ is not very realistic.
For the reconnection model, we find that the expected neutrino fluence is typically
lower than those for the fireball model by a  factor~$\sim$$5$. This results
from the fact that the pion production radius is much larger than the
\np decoupling radius, which is a very robust feature of this model. 
The condition for  inelastic \np collisions as expressed in eq.
\eqref{eq:sigmapi}, on the other hand, is fulfilled in a large range of
the parameters of the model.
It is therefore expected that \np decoupling occurs in a large fraction of GRBs
for the reconnection model.

\subsection{Gamma rays}

Secondary $\gamma$-rays resulting from \np collisions are
reprocessed by the flow due to interactions with the soft photon field
(see section~\ref{sect:gammarepr}).
This results in pair cascades for fireballs and in electron synchrotron emission for AC outflows.
The total energy (in the frame of the progenitor)
that is injected in the flow in the form of $\gamma$-rays is equal to
\beqs
\label{eq:totEgamma}
E_{\gamma} =  \frac{\Gamma_{\rm p}  \Gamma''_{\rm p} N_{\rm n} \tau
  \left< \ep''_\pi \right> \ep_{\gamma}^0}{m_\pi c^2} 
= \gamma \Gamma_{\rm np} N_{\rm n} \ep_\gamma^0 \, ,
\eeqs
which defines the factor $\gamma$. We find that $\gamma \simeq 0.5$ for both the fireball model
(for $\eta \sim$ few $\eta_\pi$) and the AC model (for $\sigma_0 \sim$  few $\sigma_{0,\pi}$). For reference
values of the parameters this implies that the fraction of the burst energy that is converted to
$\gamma$-rays is roughly $\EE{5}{-3}$ for fireballs and roughly  $\EE{2}{-3}$ for the AC model.
We assume that the bulk of the energy given in eq. \eqref{eq:totEgamma} leaves
the source after reprocessing, albeit in photons of lower energies.

\begin{figure}
\begin{center}
\includegraphics[height=8cm, angle=270]{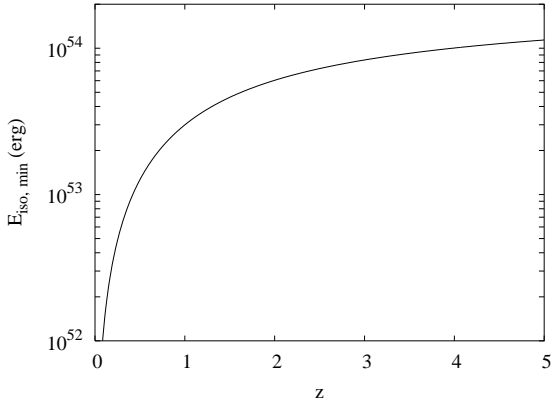}
\caption{Minimum value for the total isotropic burst energy $E_{\rm iso}$ for which the
reprocessed $\gamma$-ray emission (in the fireball model) is above the GLAST threshold, as a function of redshift.
In this figure we have taken $\xi_0 = r_{0,7} = \eta_3 = 1$, and we have taken the
burst duration equal to 10 s.
\label{fig:Eiso}}
\end{center}
\end{figure}
In section~\ref{sec:paircascades} we estimated that the $\gamma$-ray emission from pair cascades
in the fireball model is in the range $2 - 20$ GeV in the frame of the progenitor. From this we 
estimate the $\gamma$-ray number fluence $\Phi_{\gamma}$ from a source at proper distance $D_p$ as
\be
\label{eq:det:flux}
\Phi_{\gamma} = \frac{E_\gamma}{4 \pi D_p^2 \epsilon_\gamma^{\rm casc}} \, ,
\ee
where $\epsilon_\gamma^{\rm casc} \simeq 10$ GeV is the average  $\gamma$-ray energy emitted
by the pair cascades. For an energetic burst at $z=0.1$ the number fluence
is $\Phi_{\gamma} \simeq 10^{-3}$ cm$^{-2}$ which can be detected with the upcoming GLAST satellite
that has an effective area $\sim$$10^4$ cm$^2$ at these energies (Gehrels \& Michelson 1999).
In fact, we find that this emission is detectable for a fairly large range of parameters.
In figure \ref{fig:Eiso} we indicate, as a function of redshift, the minimum total isotropic
burst energy for which the gamma-ray emission by this mechanism is detectable with GLAST. 
In producing
this figure we have chosen reference values for the
relevant model parameters and assumed a burst duration of 10~s.

The isotropic equivalent energy carried by the prompt emission
at $\sim$MeV energies of a typical GRB is in the range $10^{52}$ --  $10^{54}$
erg. This is only a lower limit for the isotropic equivalent energy of the
ultrarelativistic flow which may well be a factor $\sim$10
larger than the energy carried by the prompt
emission, depending on the unknown efficiency of the mechanism that
generates the prompt emission. 
Therefore the minimum energy shown in figure \ref{fig:Eiso} is not very restrictive
and we expect that this emission is detectable for a
fairly large fraction of  GRBs in which protons and neutrons decouple.
This conclusion also holds for high redshifts where the volume for GRBs to occur
is largest.

Apart from the collisions between bulk protons and neutrons considered in this work,
pions can also be created by nuclear collisions  as a result of internal shocks
in the sub-photospheric region of the flow (M\'esz\'aros \& Rees 2000). 
This mechanism   can inject $\gamma$-rays in
the flow in a different region of the GRB parameter space.

In the AC model the energy is radiated as synchrotron emission with energy of a few hundred keV
in the observer frame (see  section~\ref{sec:synchrotron}). The corresponding energy
fluence $\sim$$10^{-5}$ erg cm$^{-2}$ is lower than  the expected  prompt emission for a burst at $z=0.1$ with
the reference values adopted here and for a typical prompt emission radiative
efficiency $\simmore$$0.1$. 
This makes it very hard to disentangle this $\gamma$-ray signal from
the prompt emission.  Of course this conclusion holds as long as the energy of the 
reprocessed $\gamma$-rays is much less than the energy of the prompt emission and the 
radiative efficiency for the prompt emission is larger than  the energy fraction $\sim$$\EE{2}{-3}$ 
transferred to $\gamma$-rays by \np collisions in the AC model.
On the other hand, even though this synchrotron component is in general
  weak, it may have a substantial contribution to the prompt X-ray emission
since its flux increases with decreasing energy as $f_{\nu}\sim \nu^{-1/2}$ 
(i.e., following the characteristic slope of fast-cooling synchrotron emission).

\section{Conclusion}
\label{sec:conclusions}

In this work we have found that $\gamma$-ray emission resulting
from inelastic collisions between differentially streaming neutrons and protons
and reprocessed by the flow may be a useful diagnostic of the nature of GRB outflows.
Provided that the baryon loading of the flow is sufficiently small, a few per mille
of the burst energy is reinjected in the flow through \np collisions in both the
fireball model and in the AC model, which was used in this work as a specific model for
GRB flows that are powered by magnetic reconnection.
In the fireball model,
the injection of these $\gamma$-rays in the outflow leads to pair cascades and subsequently to the emission
of $\gamma$-rays with observer energy in the range of 2 - 20 GeV $/ (1+z)$.
 In figure \ref{fig:Eiso}, we show the minimum total isotropic burst energy, as a
function of redshift, for which this emission can be detected by GLAST.
The constraint on the energy is not very restrictive and hence this $\gamma$-ray emission should be detectable for a
fairly large fraction of the GRBs in which \np decoupling occurs. 
In the AC model, synchrotron
energy loss prevents pair cascading and the energy is radiated away at much lower observer energies
of a few hundred keV. This component is expected to be 
dominated by the prompt $\gamma$-ray emission.

The neutrino particle fluence from  $\pi^\pm$ decay created in inelastic \np collisions in the
fireball model is found to be an order of magnitude smaller than previous
estimates. This is due to the more accurate cross sections  for elastic
and inelastic \np scattering used in this work and the distinction
between \np decoupling radius and the pion production radius.
The neutrino fluence in the AC model is smaller by another factor $\sim$$5$ due to the very
gradual acceleration of the flow, which is a very robust feature of the model.
The energy of neutrinos from \np interactions in GRB outflows as 
observed on earth is in the range 50-70 GeV 
for reference values of the parameters, which is somewhat
higher than previous estimates. We find that the observed neutrino energy in the AC model is
higher than in the fireball model because the strong magnetic field 
causes the charged pions to isotropize in the proton rest frame rather than in the collision
COM frame. Unfortunately the neutrino emission in both models is so low
that it  is very difficult to use  its properties 
to constrain the physics of GRB outflows.

In both the fireball model and the AC model we find that inelastic \np collisions occur only if the baryon
loading is sufficiently low (see section~\ref{sec:pionradius}).
For the fireball model, this condition is quite restrictive and we expect that inelastic
\np collisions are only possible for exceptional bursts.
On the other hand, inelastic collisions occur for a large range of the
parameters in the reconnection model.

The above results rely on a proper understanding of the dynamics of the flow.
We have discussed the effect of neutrons on the dynamics of the flow in
section~\ref{dynamics} (some numerical results are presented in Figs. \ref{FB1}-\ref{AC2}).
We present a numerical model which includes the acceleration of the protons due
to energy conversion in the flow, coupling of neutrons to protons by nuclear scattering
(and the dynamical decoupling of neutrons and protons) and neutron decay.
To a first approximation the dynamics of protons and neutrons can be described by the 
analytical model given in eqs. \eqref{Gammapn_simp}. This model provides a useful estimate
for the \np inelastic optical depth (section~\ref{sec:tau}) and the energies
of neutrinos (section~\ref{sec:obsnuenergy}) and $\gamma$-rays (section~\ref{sect:gammarepr}).
These estimates are generally in good agreement with results obtained from the
numerical  model described in section~\ref{dynamics} (some differences are discussed
in the main text).
The analytical estimates can be extended in a straightforward manner
to any flow with $\Gamma_{\rm p} \propto r^p$ and $\Gamma_{\rm n} =$ const.

From an observational point of view, the most promising conclusion of this work is that
$\gamma$-ray emission resulting from \np interactions may provide a signature of the nature
of the flow (section~\ref{sect:gammarepr}). The difference in energy of the reprocessed $\gamma$-ray emission between
the fireball model and the AC model results essentially from the difference in the ratio
of magnetic energy density to radiation energy density. Therefore the energy of 
this emission appears to be a robust probe for the physics of GRB outflows.
In this work we have estimated the $\gamma$-ray energy and fluence for reference values
of the burst parameters. A more detailed analysis is necessary to study the spectral properties
of the emission and compare it with other emission mechanisms
over a broad range of parameters.

It was pointed out recently that a substantial neutron component in GRB flows
may affect the properties of GRB afterglows (Belobororov 2003a). This provides a
way of constraining  the physics of GRB outflows from afterglow observations.
The numerical model discussed in this work can be used to study this possibility
in more detail. Another interesting question is whether
inhomogeneities in  the flow can cause significant particle production through \np collision
in the AC model (for fireballs, this was discussed by M\'esz\'aros \& Rees (2000)). 
These issues are left for future work.

\begin{acknowledgements}
H.K. wishes to thank the Max Planck Institute for Astrophysics, where
part of this work was completed, for its hospitality. H.K. acknowledges many valuable discussions
with Ralph Wijers and Asaf Pe'er. D.G. thanks the Astronomical Institute `Anton Pannekoek' for 
its hospitality during the initiation of this work.
We thank the referee for useful comments that helped to improve the quality
of this paper.

\end{acknowledgements}

\appendix

\section{Cross section approximations}
\label{sec:app:sigma}
\begin{figure}
\begin{center}
\includegraphics[height=8cm, angle=270]{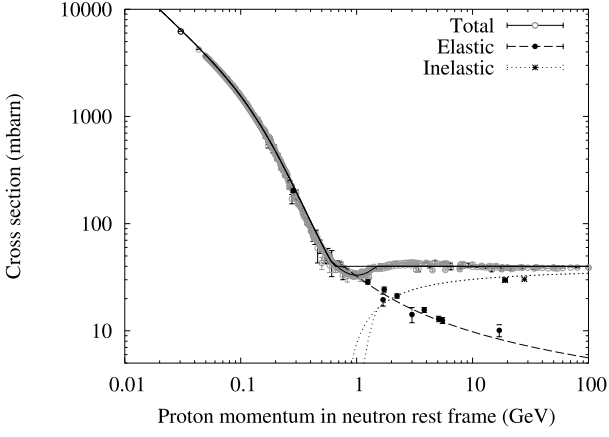}
\caption{Experimental data and approximations of the total, elastic and
inelastic \np cross sections. The thin 
lines show the approximations used for the analytical model; the thick lines show
those used in the numerical computation. \label{fig:sigma}}
\end{center}
\end{figure}
In this work we use the following approximations for the total
and elastic \np cross sections:
\be
\label{eq:app:sigma1a}
\sigma_{\ttot} =  \mathrm{max} \left[\frac{\bar{\sigma}}{0.19 \beta_{\trel} + 5.2\beta_{\trel}^3}, \bar{\sigma} \right] \, ; 
\ee
\be
\label{eq:app:sigma1b}
\sigma_{\tel} (\chi \geq \chi_\pi) =   \frac{0.75 \bar{\sigma}}{\ln \chi} \, ,
\ee
where $\bar{\sigma} \equiv \EE{4}{-26} $ cm$^2$ and $\chi \equiv \Gamma_{\rm p} / \Gamma_{\rm n}$.
At energies below the pion production threshold ($\chi < \chi_\pi$) the elastic
cross section $\sigma_{\tel} = \sigma_{\ttot}$.
Hence the inelastic cross section above the pion
production threshold can be approximated with:
\be
\label{eq:app:sigma2}
\sigma_{\tinel} (\chi \geq \chi_\pi) = \sigma_{\ttot} - \sigma_{\tel} =
\bar{\sigma} \lb 1 - \frac{0.75}{\ln \chi }\rb \, .
\ee
In these equations, $\beta_{\trel}$ and $\chi$ are related to the 
incident proton momentum in the neutron rest frame $p_{\rm p}'$  as
follows:
\be
\beta_{\trel}  \equiv  \frac{p_{\rm p}'}{\sqrt{ {p_{\rm p}'}^2 + m^2 c^2}} \, ; 
\ee
\be
\chi  \equiv \frac{\Gamma_{\rm p} }{ \Gamma_{\rm n}} = \frac{p_{\rm p}'}{m c} + \sqrt{\frac{{p_{\rm p}'}^2}{m^2 c^2} +1 } \, .
\ee
The approximations given in eqs. \eqref{eq:app:sigma1a}, \eqref{eq:app:sigma1b}  and
\eqref{eq:app:sigma2} are shown in figure \ref{fig:sigma}, together
with experimental data (Yao et al. 2006) and the approximation that
was used to describe the inelastic cross section in the numerical analysis.

\section{Frames and Lorentz factors}
\label{app:frames}
The Lorentz factor of protons and neutrons in the observer frame $K$ are denoted with $\Gamma_{\rm p}$
and $\Gamma_{\rm n}$, respectively, and we assume that both $\Gamma_{\rm p} \gg 1 $ and $\Gamma_{\rm n} \gg 1$. 
In the observer
frame, the COM frame $K''$ of the \np collision is moving with Lorentz factor
\be
\Gamma_{\rm COM} = \sqrt{\Gamma_{\rm p} \Gamma_{\rm n}} \, .
\ee
In the COM frame, protons and neutrons are moving in opposite directions with Lorentz factors
\be
\Gamma_{\rm p}'' = \Gamma_{\rm n}'' = \frac{\sqrt{s}}{2 m c^2}
= \frac{1}{2}  \lb  \frac{\Gamma_{\rm p}}{\Gamma_{\rm n}} \rb^{1/2} + \frac{1}{2} \lb \frac{\Gamma_{\rm n}}{\Gamma_{\rm p}} \rb^{1/2}  \, ,
\ee
where we take the proton and neutron masses equal to $m$. In the main text we use $K'$ to denote the
rest frame of either the proton or the neutron. If $K'$ denotes the proton rest frame, $\Gamma_{\rm p}' = 1$ by definition
and
\be
\Gamma_{\rm n}' = \frac{1}{2}  \lb  \frac{\Gamma_{\rm p}}{\Gamma_{\rm n}} +  \frac{\Gamma_{\rm n}}{\Gamma_{\rm p}} \rb
= 2 \lb \Gamma_{\rm n}'' \rb^2 -1 \, .
\ee

\end{document}